\documentclass[10pt,a4paper]{article}
\pdfoutput=1
\usepackage{jheppub}

\usepackage{amsmath,amssymb,epsfig,bm,amsfonts,yfonts,bbm,mathrsfs}
\title{Variational principle for theories with dissipation from analytic continuation}
\author{Stefan Floerchinger}
\affiliation{Institut f\"ur Theoretische Physik, Philosophenweg  16, D-69120 Heidelberg, Germany}
\emailAdd{stefan.floerchinger@thphys.uni-heidelberg.de}
\abstract{The analytic continuation from the Euclidean domain to real space of the one-particle irreducible quantum effective action is discussed in the context of generalized local equilibrium states. Discontinuous terms associated with dissipative behavior are parametrized in terms of a conveniently defined sign operator. A generalized variational principle is then formulated, which allows to obtain causal and real dissipative equations of motion from the analytically continued quantum effective action. Differential equations derived from the implications of general covariance determine the space-time evolution of the temperature and fluid velocity fields and allow for a discussion of entropy production including a local form of the second law of thermodynamics.}

\begin{document}
\maketitle

\section{Introduction}
\label{sec:Introduction}

Methods of quantum field theory are widely used to solve problems of elementary particle physics, nuclear physics, condensed matter physics or cosmology. This includes problems involving a few particles or excitations around the conventional vacuum but also thermal equilibrium states or arbitrary out-of-equilibrium dynamics. Problems of the latter type are usually treated in terms of the Schwinger-Keldysh closed time path \cite{Schwinger, Keldysh}, see for example \cite{KadanoffBaym, Danielewicz:1982kk, Chou:1984es, Calzetta:1986cq, Berges:2004yj, Rammer:2007zz, Kamenev} for introductions and overviews. This formalism is particularly useful when a coupling parameter is small such that perturbative methods can be used. Theoretical techniques that are often employed are the mapping to an (effective) kinetic theory description or resummation schemes such as the two-particle irreducible one. 

An advantage of the theoretical methods based on the closed time path is that they usually provide quite detailed microscopic information and that they are capable to follow the time evolution of arbitrary density matrices out-of-equilibrium, at least in principle. There are also no restrictions concerning the time and length scales one can consider, apart from those that are set by the range of validity of perturbative expansion schemes or other additional approximations (for example the classical statistical approximation which is sometimes used for problems involving large occupation numbers of bosonic particles). On the other side, this implies also a rather high complexity of the formalism and oftentimes high numerical effort for the concrete solution of a given problem.

For some theoretical problems, one is actually not interested in the full microscopic dynamics but rather in the behavior of the theory for large length and time scales. If it is not hindered by conservation laws or integrable dynamics, the long-time limit is usually dominated by thermalization and the corresponding equilibrium states. The thermal states have lost all memory of the initial state except for the conserved charges, in particular energy and momentum which determine the temperature and fluid velocity and conserved particle numbers which determine corresponding chemical potentials. The theoretic description of thermal states is quite universal as it only involves the thermodynamic equation of state and the standard thermodynamic formalism.

In addition to the thermal equilibrium states, also the long time dynamics is actually governed by the conservation laws to a large extend. Intuitively speaking, those modes that are fully or approximately preserved by conservation laws from relaxing to some equilibrium configuration dominate the long-time dynamics, while other modes relax on shorter time scales \cite{KadanoffMartin, Hohenberg:1977ym}. The traditional formalism to describe this is the one of hydrodynamics, or fluid dynamics in modern terminology. In addition to the conserved currents, such as the energy momentum tensor or conserved number currents, also long-range fields and order parameters such as e.\ g.\ the electromagnetic field or the magnetization need to be included in the theoretical description \cite{Hohenberg:1977ym}. 

The fluid dynamic description is not quite as universal as thermodynamics because it needs additional information about transport properties, for example the shear and bulk viscosity or dissipative damping terms for the long range fields and order parameters. The range of applicability of a fluid dynamic description is set by the time (and associated length) scale of thermalization. The thermalization time is shorter, and therefore the range of applicability of a fluid dynamic description larger, when interaction effects are stronger.

It is possible to obtain fluid dynamics as the long time limit of kinetic theory. Theoretical methods such as the Chapman-Enskog expansion \cite{ChapmanCowling} or Grad's method of moments (see e.\ g.\ \cite{Denicol:2012es, Denicol:2012cn}) describe the mapping in detail. In terms of these methods one can also determine the transport properties by microscopic calculations based on kinetic theory. While this approach is fine for the situations where it is applicable (such as dilute gases) it has the conceptual disadvantage that it needs both a weak coupling assumption for the applicability of kinetic theory and a form of a strong coupling assumption for the applicability of a fluid dynamic description. One may therefore wonder whether there exists a shortcut in the theoretical setup that leads more directly from the formulation of a quantum field theory to a fluid dynamic type of formalism. Such a ``shortcut formalism'' should be capable of describing local equilibrium situations and the corresponding dynamics but without relying on the weak coupling assumptions underlying kinetic theory. For thermodynamic equilibrium, such a formalism exists of course in the form of the imaginary time or Matsubara formalism and it is heavily used also for non-perturbative calculations (e.\ g.\ in the setup of lattice gauge theory). 

It is plausible that there should also be a general quantum field theoretic formalism which is applicable out-of-equilibrium, however not in the sense of arbitrary, far-from-equilibrium states but rather for states that can be described by an approximate {\it local} equilibrium. Such a situation may be coined as close-to-equilibrium. One formalism of this type is linear response theory. It is restricted to small excitations around global equilibrium states but within this limitation it allows for non-perturbative statements (in the sense of the expansion in a coupling constant), for example the fluctuation-dissipation theorem or the Green-Kubo relations. 
One may also argue that for strongly interacting theories with a gravitational dual, such a close-to-equilibrium formalism is used already at least implicitly in many calculations based on the AdS/CFT correspondence. 

It should then also be possible to construct such a close-to-equilibrium formalism using solely the methods of quantum field theory without relying on perturbative assumptions and the mapping to kinetic theory. These considerations are part of the motivation for the present work.

The formalism we aim to develop here is a direct generalization of equilibrium quantum field theory. Instead of using the full machinery of the Schwinger-Keldysh closed time path, we will use analytic continuation\footnote{The analytic continuation technique has a long history as an approach to non-equilibrium problems, see for example refs.\ \cite{KadanoffMartin, Abrikosov}.}. This has the advantage that part of the formal complexity of the closed time path formalism can be avoided. For example, we will not need a full doubling of the fields corresponding to the two branches of the time contour. However, as will be discussed in detail below, we will have to introduce a specific sign operator that allows to parametrize the branch cuts in the analytic (generalized) frequency plane associated with dissipative behavior. Because the sign operator allows to distinguish between the two sides of the branch cut, and therefore retarded and advanced correlation functions, a few aspects of the resulting formalism work in practice similar as for the closed time path formalism. In particular, and that is the main result of this paper, it is thereby possible to obtain real and causal equations of motion including dissipative effects for field expectation values from the variation of an analytically continued version of the one-particle irreducible quantum effective action.

The fact that our formalism is based on an effective action implies that not only the field equations or equations of motion can be studied, but also very directly various correlation functions which follow from functional derivatives of the analytic action. This is in contrast to the traditional formulation of fluid dynamics which starts directly with the equations of motion. The correlation functions obtained in this way describe generalizations of thermal equilibrium fluctuations in various fields. In ref.\ \cite{Landau:1980mil}, such fluctuations are called quasi-stationary. 

The analytic continuation we will study starts from an Euclidean description of {\it local} equilibrium states. In the presence of additional field expectation values or order parameters, these local equilibrium states should be understood in the sense of generalized Gibbs ensembles \cite{Hayata:2015lga}. One should also keep in mind that local equilibrium is typically only an approximation to the full dynamics (although it can be a rather good approximation). The formalism assumes implicitly that the theory is probed on length and time scales that are large compared to the thermalization dynamics, or, in other words, that approximate local thermalization is efficient enough for those degrees of freedom that make up the bath. There are, of course, also situations where this assumption is not justified. 

We do not assume, however, that dissipative processes are absent. In contrast, their proper description will be one of the main points of the present paper. A particularly nice feature of the close-to-equilibrium formalism is that it allows a straight-forward discussion of entropy production. This includes entropy production due to dissipative processes concerning a (quantum) field expectation value $\Phi_a$ as well as shear and bulk viscous dissipation. 

As will be discussed in detail below, the temperature and fluid velocity do not have the same status as quantum field expectation values or order parameter fields $\Phi_a$. In particular, the analytic effective action is not stationary with respect to variations of the temperature or fluid velocity. These fields should rather be seen as certain parameter fields for which the dynamics does not directly follow from the equations of motion but rather somewhat implicitly from general coordinate invariance or the closely connected conservation laws for energy and momentum. More details are discussed below.

The formalism presented here allows to derive the equations of motion of fluid dynamics from the variation of an effective action. This is discussed for a fluid without conserved charges apart from energy and momentum within the so-called first order approximation in section \ref{sec:FluidDynamics}. The extension to more general fluids as well as the second order formalism will be discussed elsewhere. Attempts to construct a variational principle that yields the equations of fluid dynamics have a long history, see for example \cite{Andersson:2006nr} for an overview. Recently, fluid dynamics was discussed extensively in the context of effective field theory \cite{Dubovsky:2005xd, Nicolis:2011ey, Dubovsky:2011sj, Dubovsky:2011sk, Nicolis:2011cs, Bhattacharya:2012zx, Endlich:2012vt, Grozdanov:2013dba, Andersson:2013jga, Nicolis:2013lma, Haehl:2013hoa, Ballesteros:2014sxa, Geracie:2014iva, Delacretaz:2014jka, Kovtun:2014hpa, Haehl:2014zda, Haehl:2015pja, Harder:2015nxa, Crossley:2015evo}.  While actions for ideal fluid dynamics can be constructed in different ways, it is more difficult to treat dissipation, for a recent discussion see \cite{Crossley:2015evo}. 

The formalism developed here differs from earlier works in several ways. Most prominently, it is based on the analytic continuation of the one-particle irreducible effective action. The effect of quantum and thermal fluctuations or noise is already included in this object. No functional integrals over fluid variables are part of this formalism (except possibly to describe additional initial state fluctuations). The effect of quantum and thermal fluctuations is taken into account by the functional integrals that define the quantum effective action in the Euclidean domain, i.\ e.\ before analytic continuation. 

This setup implies in particular that the dissipative equations of motion derived from the analytic effective action for field expectation values $\Phi_a$ are {\it not} subject to further renormalization effects by quantum and thermal fluctuations. In the presence of initial state fluctuations in the fluid fields, another level of statistical description is added which may of course be described by correspondingly defined expectation values and correlation functions as well as renormalized equations of motion, see for example \cite{Blas:2015tla}. 

The one-particle irreducible effective action is very closely related to the partition function in the grand canonical ensemble. It was recently discussed how one can obtain constraints for the non-dissipative terms of fluid dynamics from investigations of the partition function \cite{Banerjee:2012iz, Jensen:2012jh, Bhattacharyya:2013lha, Bhattacharyya:2014bha, Becattini:2015tpl}. The formalism developed here could be used to generalize these considerations such that also dissipation is taken into account. 

The present paper is organized as follows. We start in section \ref{sec:DampedHO} with an introductory discussion of the simple, damped harmonic oscillator. Some of the main elements of our the formalism can already be introduced there. In section \ref{sec:AnalyticQuantumEffectiveAction} we turn then to quantum field theory. In particular, the quantum effective action for a situation with locally varying temperature and fluid velocity is defined first in the Euclidean domain in subsection \ref{sec:GeneratingFunctionalsEuclidean}. In order to understand properly the analytic structure of the quantum effective action, we discuss the analytic structure of two-point functions in subsection \ref{sec:ACTwoPoint} and of more general correlation functions in subsection \ref{sec:ACHigherOrderCF}. The general structure of the analytically continued quantum effective action (or analytic effective action for short) is discussed in subsection \ref{sec:GeneralStructureAnalyticEffectiveAction} and the relation to other actions, such as the often considered time-ordered or Feynman effective action is considered in subsection \ref{sec:RelationToTimeOrderedMatsubara}. 

In section \ref{sec:GeneralizedVariationalPrinciple} we discuss a generalized variational principle that allows to obtain dissipative equations of motion directly from the analytic effective action. We discuss there also the concept of a retarded functional derivative and why this leads to causal equations of motion. In section \ref{sec:EnergyMomentumConservationEntropyProduction} we discuss the important issue of energy momentum conservation as well as entropy production. More specific, subsection \ref{sec:EnMomTensor} discusses how one can obtain the expectation value of the energy-momentum tensor from the analytic effective action, and subsection \ref{sec:EnergyMomentumConservation} discusses general coordinate covariance and the closely related issue of energy-momentum conservation. We obtain there differential equations that can be used to fix the space-time evolution of the temperature and fluid velocity. Particularly interesting is an equation one can derive in that way for entropy production. This is discussed in subsection \ref{sec:Entropy} and also a local form of the second law of thermodynamic is stated there. 

The different theoretical concepts and equations are then discussed for a particular example of an effective action that describes a scalar field with $O(N)$ symmetry in section \ref{eq:ExampleScalarField} as well as for an action describing viscous relativistic fluid dynamics in section \ref{sec:FluidDynamics}. Finally, some conclusions are drawn in section \ref{sec:Conclusions}. Appendix \ref{sec:appA} contains a compilation of useful relations about various two-point correlation functions as they can be derived via linear response theory, for example the fluctuation-dissipation relation, Onsager's relations or the spectral representation. 

\section{The damped harmonic oscillator}
\label{sec:DampedHO}

The equation of motion of a simple, damped harmonic oscillator is
\begin{equation}
m \ddot x = - k x - c \dot x,
\end{equation}
which can also we written as
\begin{equation}
\ddot x + 2 \zeta \omega_0 \dot x + \omega_0^2 x = 0,
\label{eq:EquationOfMotionDamped}
\end{equation}
with the undamped frequency $\omega_0 = \sqrt{k/m}$ and the so-called damping ratio $\zeta = c/\sqrt{4 m k}$. Without the damping term one can obtain the equation of motion from the variation of the action
\begin{equation}
S[x] = \int d t  \left\{ \frac{1}{2}m \dot x^2 - \frac{1}{2} k x^2 \right\} = \int \frac{d \omega}{2\pi} \; \frac{m}{2} x^*(\omega) \left[ \omega^2 - \omega_0^2 \right] x(\omega).
\label{eq:ActionUndamped}
\end{equation}
In the last equation we introduced the Fourier transform
\begin{equation}
x(t) = \int \frac{d \omega}{2\pi} e^{-i \omega t} \; x(\omega) ,
\end{equation}
and from $x(t) \in \mathbb{R}$ it follows that $x^*(\omega) = x(-\omega)$.

To find an action for the damped harmonic oscillator one might be tempted to replace simply the kernel $\left[ \omega^2 - \omega_0^2 \right]$ on the right hand side of \eqref{eq:ActionUndamped} by $\left[ \omega^2 + 2 i \omega \zeta \omega_0 - \omega_0^2 \right]$. However, the additional term linear in $\omega$ would simply cancel out -- it is odd with respect to $\omega \to - \omega$ while $x^*(\omega) x(\omega)$ is even. In the time representation such a term would be a total derivative and would therefore not contribute to the field equations.

The kernel of the Fourier representation on the right hand side of eq.\ \eqref{eq:ActionUndamped} plays the role of the inverse propagator. It has zero-crossings at $\omega=\pm \omega_0$ corresponding to the two poles of the propagator or the two independent solutions of the equation of motion \eqref{eq:EquationOfMotionDamped} for $\zeta = 0$. The general expectation for an effective propagator in the presence of dissipative mechanisms is that poles are broadened and become branch cuts. For the inverse propagator this implies that zero-crossings are avoided by additional discontinuous terms. Consider for example the inverse propagator
\begin{equation}
\omega^2 + 2 i s_\text{I}(\omega) \,  \omega \, \zeta \omega_0 - \omega_0^2 ,
\label{eq:inversePropagatorCut}
\end{equation}
where
\begin{equation}
s_\text{I}(\omega) = \text{sign}\left( \text{Im} \, \omega \right).
\end{equation}
Formally, the zero crossings of \eqref{eq:inversePropagatorCut} are at
\begin{equation}
\omega = - i s_\text{I}(\omega) \zeta \omega_0 \pm \sqrt{\omega_0^2 - \zeta^2 \omega_0^2} .
\end{equation}
However, this equation has no solutions because the imaginary terms on both sides always have opposite signs. In other words, for $\text{Im} \, \omega >0 $ it appears as if the zero-crossing of \eqref{eq:inversePropagatorCut} were at $\text{Im} \, \omega < 0 $ and {\it vice versa}. 

One could, however, also say that the expression in \eqref{eq:inversePropagatorCut} has several zero-crossings on different Riemann sheets. On the Riemann sheet that corresponds to the analytic continuation of the upper half of the complex $\omega$-plane, the zero-crossings are at $\omega=- i \zeta \omega_0 \pm \sqrt{\omega_0^2 - \zeta^2 \omega_0^2}$, while on the Riemann sheet that is analytically connected to the lower half of the complex $\omega$-plane they are at $\omega= i \zeta \omega_0 \pm \sqrt{\omega_0^2 - \zeta^2 \omega_0^2}$. As will be discussed below, the zero-crossings on the two Riemann sheets correspond to the dissipative equations of motion for forward and backward time evolution.

One may now try for the {\it effective} action of the damped harmonic oscillator the expression (we use conventions where the effective action $\Gamma[x]$ differs from $S[x]$ in overall sign)
\begin{equation}
\begin{split}
\Gamma[x]  = & \int \frac{d \omega}{2\pi} \; \frac{m}{2} x^*(\omega) \left[ - \omega^2 - 2 i \, s_\text{I}(\omega)\,  \omega \, \zeta \omega_0 + \omega_0^2 \right] x(\omega) \\
 = & \int d t  \left\{ - \frac{1}{2}m \dot x^2 + \frac{1}{2} c\,  x \, s_\text{R}(\partial_t) \dot x + \frac{1}{2} k x^2 \right\} .
\end{split}
\label{eq:actionDampedHarmonicOscillator}
\end{equation}
In the second line we have gone back to the time representation and replaced
\begin{equation}
s_\text{I}(\omega) = \text{sign}( \text{Im} \, \omega) \to \text{sign}(\text{Im} \, i \partial_t) = \text{sign}(\text{Re} \, \partial_t)= s_\text{R}(\partial_t) .
\end{equation}
Note that due to the symbol $s_\text{I}(\omega)$ in the first line and $s_\text{R}(\partial_t)$ in the second line of \eqref{eq:actionDampedHarmonicOscillator} the damping terms are formally even under the symmetry $\omega \to - \omega$ in the frequency representation and cannot be written as total derivatives in the time representation.

The variation of the action \eqref{eq:actionDampedHarmonicOscillator} gives up to boundary terms 
\begin{equation}
\begin{split}
\delta \Gamma  = & \int d t  \left\{ m \ddot x \, \delta x - c\,  \dot x \, s_\text{R}(\partial_t) \delta x +  k x\,  \delta x \right\} .
\end{split}
\label{eq:variationActionDampedOscillatorTime}
\end{equation}
We have used here that $s_\text{R}(\partial_t)$ is an odd function in $\partial_t$ in order to perform partial integration. One sees from \eqref{eq:variationActionDampedOscillatorTime} that the principle of stationary action $\delta \Gamma = 0$ leads to the right equation of motion for forward time propagation if one demands that $s_\text{R}(\partial_t) \delta x = - \delta x$. Similarly, the principle of stationary action for variations with $s_\text{R}(\partial_t) \delta x = \delta x$ leads to the equation of motion with reversed time direction.
One could also write the variation of the action as
\begin{equation}
\delta \Gamma = \int d t  \left\{ \delta x  \, m \ddot x + \delta x \, c\,   s_\text{R}(\partial_t)\dot x  + \delta x \, k x \right\} ,
\label{eq:variationActionDampedOscillatorTime2}
\end{equation}
which gives the correct equation of motion when one sets $s_\text{R}(\partial_t) \dot x = \dot x$ for forward time evolution and $s_\text{R}(\partial_t) \dot x = - \dot x$ for backward time evolution after the variation. More general, for forward time evolution, one has to set $s_\text{R}(\partial_t) \to -1$ if $\delta x$ is to the right of this operator and $s_\text{R}(\partial_t)\to 1$ if $\delta x$ is to the left of the operator $s_\text{R}(\partial_t)$. This will be discussed in more detail below.

One can also obtain the equations of motion for forward or backward time evolution, respectively, from variation of the action \eqref{eq:actionDampedHarmonicOscillator} in frequency representation. A summary of the correct variations to obtain the dissipative equations of motion is given in table \ref{tab:variations}.
\begin{table}
\begin{tabular}{ l | l | l | l }
  \hline \hline                      
  forward time ev. & $s_\text{R}(\partial_t) \delta x(t) = - \delta x(t)$ &
   $\delta x(\omega)$ with $s_\text{I}(\omega) = - 1$ & 
   $\delta x^*(\omega)$ with $s_\text{I}(\omega) = + 1$ \\ \hline
  backward time ev. & $s_\text{R}(\partial_t) \delta x(t) = + \delta x(t)$ & 
  $\delta x(\omega)$ with $s_\text{I}(\omega) = + 1$ & 
  $\delta x^*(\omega)$ with $s_\text{I}(\omega) = - 1$ \\ \hline \hline
\end{tabular}
\caption{Variations for which the principle of stationary action $\delta \Gamma=0$ leads to the correct dissipative equations of motion with forward and backward time evolution, respectively.}
\label{tab:variations}
\end{table}

It is interesting to extend the action \eqref{eq:actionDampedHarmonicOscillator} to curved space by introducing a 0+1-dimensional metric $g(t)>0$ with inverse $g^{-1}$. From general covariance, the extension of \eqref{eq:actionDampedHarmonicOscillator} is
\begin{equation}
\Gamma[x,g]= \int d t \sqrt{g} \left\{ - \frac{1}{2}m g^{-1} \dot x^2 + \frac{1}{2} c\,  x \, s_\text{R}(g^{-\frac{1}{2}} \partial_t) g^{-\frac{1}{2}} \dot x + \frac{1}{2} k x^2 + \Phi_G(T) \right\} .
\label{eq:actionHOwithMetric}
\end{equation}
We have also added here a term $\Phi_G(T)$ that accounts for the grand canonical potential of the bath degrees of freedom that are not described explicitly by the action \eqref{eq:actionHOwithMetric} but have been ``integrated out'' already\footnote{To see that this is indeed the correct expression one considers the time integral along the Matsubara contour from some time $t$ to $t-i\beta^0$. In general, local thermal equilibrium is described by a manifold with periodicity in complex space-time coordinates such that bosonic (fermionic) fields satisfy $\phi(x^\mu)=\pm\phi(x^\mu-i\beta^\mu)$ where the inverse temperature vector $\beta^\mu=u^\mu/T$ is determined by the fluid velocity $u^\mu$ and the temperature $T$.}. 

One can actually use partial integration to bring \eqref{eq:actionHOwithMetric} to the alternative form 
\begin{equation}
\Gamma[x,g]= \int d t \sqrt{g} \left\{ - \frac{1}{2}m g^{-1} \dot x^2 - \frac{1}{2} c\,  g^{-\frac{1}{2}} \dot x \, s_\text{R}(g^{-\frac{1}{2}} \partial_t) x  + \frac{1}{2} k x^2 + \Phi_G(T) \right\} .
\label{eq:actionHOwithMetric2}
\end{equation}
For this one considers $s_\text{R}(\cdot)$ to be an odd function of the argument and shows that a similar partial integration as above is actually possible for any odd function. 

From the variation of $\Gamma$ with respect to $g$ one can obtain the expectation value of the 0+1-dimensional analog of the energy-momentum tensor, 
\begin{equation}
\frac{1}{2}\sqrt{g} \, T^{00} = \frac{\delta \Gamma}{\delta g} .
\end{equation}
The dissipative term does not contribute to this; the formal reason will be discussed below. For the variation of the grand canonical potential term one has to write $T=1/(\sqrt{g} \beta^0)$ and keep $\beta^0$ fixed. Generally, in local thermal equilibrium, the temperature $T$ is related to the inverse temperature vector $\beta^\mu = u^\mu/T$ or $T=1/\sqrt{-g_{\mu\nu}\beta^\mu \beta^\nu}$ and $\beta^\mu$ should be kept fixed when the metric is varied. The inverse temperature $\beta^\mu$ does not play the role of a dynamical field but should rather be seen as an external parameter field which determines the local equilibrium state. It's time dependence is fixed by the conservation laws of energy and momentum.

Using $d\Phi_G=-S dT$ and $\Phi_G = U - T S$, where $U$ is the thermal energy of the bath variables, gives for the total energy associated with \eqref{eq:actionHOwithMetric}  (setting now $g=1$)
\begin{equation}
E = T^{00} = \frac{1}{2} m \dot x^2 +\frac{1}{2} kx^2 + U(T).
\label{eq:energyHO}
\end{equation}
Although in the presence of viscous damping the mechanical energy and the thermal energy are not conserved separately, their sum is. Within a local equilibrium assumption one can actually use the energy conservation law $dE/dt = 0$ to determine the time-dependence of the temperature $T(t)$ from the time dependence of the mechanical energy which, in turn, follows via the dissipative equations of motion.

\section{The analytic quantum effective action}
\label{sec:AnalyticQuantumEffectiveAction}

After the introductory discussion of the simple damped harmonic oscillator, let us now turn to field theory. For quantum field theories that respect the principles of unitarity, dissipative effects are not present at the level of the fundamental theory described by the microscopic action. However, they can arise in specific sense on the level of the quantum effective action. This will be discussed in detail below. To gain some intuitive understanding how it is possible that effective dissipative terms arise in the transition from the microscopic action to the quantum effective action consider the following two examples.

The first example is the decay of an unstable particle or resonance. For concreteness take a muon that can decay into an electron and two neutrinos. The theory is unitary on the microscopic level, but the decay width appearing in the effective muon propagator can also be seen as a dissipative property.  One can formally ``integrate out'' the fields corresponding to the decay products. This corresponds to a partial trace over the corresponding part of the full density matrix. In the remaining sector describing the muon, the dynamics is described by a reduced density matrix and can appear as dissipative. In particular, energy and momentum do not have to be conserved there. The situation is similar for strong electromagnetic fields with field strengths above the Schwinger threshold of electron-positron pair production. In a description where electrons and positrons are ``integrated out'' or ``traced out'', this appears as a dissipative effect.

The second example is a thermal situation where many degrees of freedom play a role such that a complete microscopic description becomes intractable. A statistical description at non-zero temperature leads to modifications of the quantum effective action (and the corresponding equations of motion) and in particular additional dissipative terms can appear. Energy and information can be transferred to the heat bath and lead to an increase of the thermal energy and entropy.

In the following we will develop a theoretical formalism that can account for the physics of these effects which is based on the analytically continued quantum effective action, or analytic effective action for short. Alternative formulations could be based on the Schwinger-Keldysh closed time path effective action, see for example \cite{Calzetta:1986cq}. The two formalisms differ somewhat in practical terms. They also address different types of problems. While the closed time path effective action allows to address general far-from-equilibrium situations, the formalism discussed below is a direct generalization of the equilibrium formalism and therefore more suited for close-to-equilibrium situations.

\subsection{Generating functionals in the Euclidean domain}
\label{sec:GeneratingFunctionalsEuclidean}

We start the constructions by discussing different generating functionals and in particular the quantum effective action in Euclidean space. It can then be analytically continued in the subsequent step. One may take as a starting point the partition function of a theory described in terms of fundamental fields $\phi_a(x)$ in the functional integral representation. In principle, $\phi_a$ might have bosonic and fermionic components, the latter being represented by Grassmann fields. For the present study we shall assume at some places bosonic fields for simplicity but the generalization to Grassmann fields is straight forward with some additional care concerning minus signs from the interchanging of fields.

In the presence of corresponding source terms $J_a(x)$, the partition function is
\begin{equation}
Z[J] = \int D \phi \; e^{-S_E[\phi] +\int_x J \phi}.
\label{eq:PartFunct}
\end{equation}
The configuration space for the Euclidean action $S_E[\phi]$ depends on the vacuum state of the theory. We will be interested here in (approximate) {\it local} equilibrium states that can be described by a temperature $T(x)$ and a fluid velocity $u^\mu(x)$. They will enter the construction in terms of the combination $\beta^\mu=u^\mu/T$. For convenience we will use a general coordinate system with metric $g_{\mu\nu}(x)$. Global thermal equilibrium is included as a special case and in particular also the conventional vacuum with vanishing temperature $T\to 0$.

From the similarity between the local density matrix (we use metric with signature (-,+,+,+))
\begin{equation}
e^{\beta^\mu(x) \mathscr{P}_\mu}
\end{equation}
with the translation operator 
\begin{equation}
e^{i \Delta x^\mu \mathscr{P}_\mu}
\end{equation}
one is lead to the representation of the partition function $Z[J]$ in terms of a functional integral on a geometry with periodicity in imaginary space direction such that bosonic (fermionic) fields satisfy $\phi(x^\mu-i\beta^\mu(x)) = \pm \phi(x^\mu)$. 

The microscopic Euclidean action that enters \eqref{eq:PartFunct} is defined by analytic continuation from the space-time with Minkowski signature. More specific, one has 
\begin{equation}
S_E[\phi] = - i S[\phi]{\Big |}_\text{analy. cont.} = -i \int d^d x \sqrt{g} \, \mathscr{L}
\label{eq:EuclideanActionLagrangian}
\end{equation}
where $\sqrt{g} = \sqrt{-\text{det} \;g_{\mu\nu}}$ is the determinant of the metric (with Minkowski signature) and $\mathscr{L}$ is the Lagrangian. We use here a formulation where the metric is not modified by the analytic continuation but the space-time differential $dx^\mu$ is complex. More specific, in a coordinate system where the fluid velocity points in the $x^0$ direction, $u^\mu=(1,0,0,0)$, one has $dx^\mu = (-i d\tau/T, d x^j)$ where $d\tau\in[0,1]$ and the $d x^j$ are real. In other coordinate systems, all components $dx^\mu$ are in general complex. 

To make this more concrete, we consider a time-like hypersurface $\Sigma$ which can be defined in terms of a scalar time function $t(x)$ as the manifold with $t(x)=\bar t$. To each position $x^\mu$ on this base manifold we associate a Matsubara circle in imaginary direction $-i\beta^\mu(x)$ and write the position on the product Manifold $\Sigma \times M$ as $x^\mu-i\tau \beta^\mu(x)$ where $\tau\in [0,1]$.
Correspondingly, we decompose the $d$ dimensional differential volume element $d^dx \sqrt{g}=d\Sigma_\mu dx_M^\mu$ into the integral over a $d-1$ dimensional hypersurface $d\Sigma_\mu$ and the Matsubara integral in the imaginary flow time direction $dx^\mu_M=-i \beta^\mu d\tau$. Indeed, if the hypersurface $\Sigma$ is defined in terms of a time function $t(x)$ as the manifold with $t(x)=\bar t$, one has
\begin{equation}
d\Sigma_\mu = d^d x \sqrt{g} \, \delta(\bar t - t(x)) \, \partial_\mu t(x) ,
\end{equation}
and for any direction $d x^\mu$ that is not orthogonal to $\partial_\mu t(x)$ one can write
\begin{equation}
d\Sigma_\mu dx^\mu = d^d x\sqrt{g} \, \delta\left( \bar t - t(x) \right) \, d x^\mu  \partial_\mu t = d^d x \sqrt{g} \, \delta\left( \bar t - t(x) \right) \, dt(x) = d^d x \sqrt{g}.
\label{eq:dSigmamudxmu}
\end{equation}
This shows that the volume element in the Euclidean domain $d^d x \sqrt{g}=d\Sigma_\mu dx^\mu_M=-i \beta^\mu d\Sigma_\mu d\tau$ contains a factor $-i$. 

The Lagrangian is of the standard form, for example for a scalar field with $\mathbbm{Z}_2$ symmetry,
\begin{equation}
- \mathscr{L} = \frac{1}{2} \partial_\mu \varphi \partial^\mu \varphi+ \frac{1}{2}m^2 \varphi^2+ \frac{1}{4!} \lambda \varphi^4.
\label{eq:LagrangianScalar}
\end{equation}
Note, however, that similar to the differential $dx^\mu$ also the derivatives are complex. In a coordinate system where $u^\mu$ points in $x^0$ direction one has $ \partial_0=\frac{\partial}{\partial x^0} = i \frac{\partial}{\partial \bar x^0}= i  \partial_{\bar 0}$ whereas the other derivatives are real. For Minkowski metric, the kinetic term in \eqref{eq:LagrangianScalar} becomes $\frac{1}{2}(\partial_{\bar 0}\varphi \partial_{\bar 0}\varphi+\partial_j\varphi\partial_j\varphi)$ which explains in which sense the analytically continued theory is indeed of Euclidean form. More general, the derivatives $\partial_\mu$ are complex.

In addition to the action $S_E[\phi]$, also the source term in \eqref{eq:PartFunct} is evaluated in the Euclidean domain. More specific, the abbreviated form in \eqref{eq:PartFunct} stands for
\begin{equation}
\int_x J \phi = i \int d^d x \sqrt{g} \left\{ J(x) \phi(x) \right\}
\label{eq:complextimeJphi}
\end{equation}
where similar as for $S_E[\phi]$ the volume integral is over the imaginary, compact direction $dx_M^\mu = -i u^\mu d\tau$ and the space-like hypersurface $\Sigma_\mu$. Functional derivatives are defined such that for example (recall that $d^dx$ contains a factor $-i$)
\begin{equation}
\frac{1}{\sqrt{g}(y)}\frac{\delta}{\delta J(y)} \int_x J \phi =  \phi(y).
\label{eq:functDeriv}
\end{equation}
We use here a $d$-dimensional functional derivative where the space-time argument $y$ in \eqref{eq:functDeriv} consists of a (real) base point and an imaginary part which parametrizes the position on the Matsubara axis, $y^\mu=\text{Re} ( y^\mu) - i \tau \beta^\mu$. Accordingly, on the right hand side of eq.\ \eqref{eq:functDeriv} the integral over the Matsubara circle is reduced to a sinlge point.

For some purposes, one actually needs a $d-1$ dimensional functional derivative that acts at a specific position on the surface $\Sigma_\mu$ but does not reduce the integral along the Matsubara circle to a single point. This is in particular needed to evaluate thermal expectation values of composite fields such as the energy-momentum tensor. In the present work, we will use the $d$-dimensional functional derivatives in such cases but add an integral along the Matsubara direction afterwards by hand. For this prescription, auxiliary fields like the metric $g_{\mu\nu}(x)$ are formally extended to complex coordinates $x^\mu-i\tau \beta^\mu(x)$, as well, although final expressions will always be evaluated for configurations where $g_{\mu\nu}$ is independent of the Matsubara time $\tau$.

To repeat, the action in the Euclidean domain $S_E[\phi]$ lives on a configuration space which associates to every base point $x^\mu$ one imaginary direction $dx_M^\mu = -i u^\mu d\tau $ which is compactified at finite temperature.  The other $d-1$ directions parametrize a space-like hypersurface $\Sigma_\mu$ of the complete space-time. The volume integral $\int d^d x \sqrt{g}$ on the right hand side of \eqref{eq:EuclideanActionLagrangian} goes along the hyper surface $\Sigma_\mu$ and once around the loop in the compactified imaginary direction. This construction can be made more explicit in terms of the ADM formalism when needed \cite{Banerjee:2012iz, Hayata:2015lga}. For the present purpose, we will not need further details. 

The positioning of the surface $\Sigma_\mu$ in time direction can be varied freely if one assumes that local equilibrium in the sense of a generalized Gibbs ensemble holds everywhere. It is usually convenient to place the surface such that it intersects with a space-time point where expectation values should be calculated. 
The local temperature $T(x)$ and fluid velocity $u^\mu(x)$, that describe a local equilibrium configuration are at this point not specified further. Ultimately, they will be fixed by general covariance and the closely related covariant conservation of energy and momentum as well as initial values. One should keep in mind that a local equilibrium picture is based on physical assumptions which hold in general at best approximately. In particular, the picture assumes that the processes that drive thermalization are efficient enough such that the local equilibrium picture can be applied on the time- and length scales one wishes to study. In a very homogeneous situation, or on very large space and time scales, this is typically the case and thermal equilibrium corresponds to the unique state of the theory for given conserved quantities, in particular energy and momentum.

It is important to emphasize that local equilibrium differs in general from a complete, global equilibrium. Deviations from the latter are parametrized by gradients of temperature and fluid velocity but also by other fields that deviate from their equilibrium configuration. In the formalism we develop, these additional, non-equilibrated fields are followed explicitly and are in this sense not part of the ``thermal bath''. In this situation one might understand the local equilibrium configuration in terms of a generalized Gibbs ensemble.

By taking functional derivatives of $Z[J]$ with respect to the sources one can obtain expectation values and correlation functions corresponding to the state described by $S_E[\phi]$. It is useful to introduce also the Schwinger functional in the Euclidean domain $W_E[J]$ by the definition
\begin{equation}
Z[J] = e^{W_E[J]}.
\label{eq:defSchwingerFunct}
\end{equation}
Repeated functional derivatives of $W[J]$ yield connected correlation functions while a single functional derivative yields the expectation value of $\phi$,
\begin{equation}
\frac{1}{\sqrt{g}(x)}\frac{\delta}{\delta J_a(x)} W_E[J]  = \Phi_a(x) = \langle \phi_a(x) \rangle .
\end{equation}
The quantum effective action is a functional of field expectation values. Formally it is defined in the Euclidean domain as the Legendre transform of $W_E[J]$,
\begin{equation}
\Gamma_E[\Phi] = \int_x J_a(x) \Phi_a (x) - W_E[J] \quad\quad\quad \text{with} \quad\quad\quad \Phi_a(x) = \frac{1}{\sqrt{g}(x)}\frac{\delta}{\delta J_a(x)} W_E[J].
\label{eq:defGamma1PI}
\end{equation}
Repeated functional derivatives of $\Gamma_E[\Phi]$ give one-particle irreducible correlation functions. This may seem to be a quite formal statement but it has very interesting implications: A complete connected correlation function can be determined by adding only tree-level expressions but with effective vertices and propagators obtained from the functional derivatives of $\Gamma_E[\Phi]$ instead to those from the functional derivatives of the microscopic action $S_E[\phi]$. 

The effective action in the Euclidean domain $\Gamma_E[\Phi]$ is also subject to the following field equation, which can be directly derived from eq.\ \eqref{eq:defGamma1PI},
\begin{equation}
\frac{\delta}{\delta \Phi_a(x)}\Gamma_E[\Phi] = \sqrt{g}(x) \, J_a(x) .
\label{eq:GammaFieldEquation}
\end{equation}
For vanishing source, $J_a(x)=0$, this equation resembles the classical equation of motion that follows from Hamilton's principle for a classical field theory, $\delta S[\phi] = 0$. However, so far the effective action $\Gamma_E[\Phi]$ is defined in the Euclidean domain where it accounts for the trace over a density matrix. Only statistical expectation values and correlation functions on a space-like hypersurface can be calculated directly from this. Note that the action $S_E[\phi]$ and similarly the Schwinger functional $W_E[J]$ and effective action $\Gamma_E[\Phi]$ in the Euclidean domain depend formally also on the choice of the space-like hypersurface $\Sigma_\mu$. For example, one can calculate correlation functions for fields at different positions on this surface from the functional derivatives of $W_E[J]$. Correlation functions between points that are not on the surface $\Sigma_\mu$ cannot be calculated, yet. 

Before we obtain a dynamical equation of motion for the field expectation value $\Phi$ from $\Gamma_E[\Phi]$ we need to study analytic continuation and the analytic structure of $\Gamma_E[\Phi]$. By the latter we mean here the analytic structure of the correlation functions that follow from the functional derivatives of $\Gamma_E[\Phi]$ as a function of the space-time arguments. The notion will become more clear in due course.

The result of this analysis will be an analytically continued effective action $\Gamma[\Phi]$ from which one can derive correlation functions or Greens functions that go beyond the expectation values in local thermal equilibrium on a fixed space-like hypersurface. In particular, the analytically continued effective action will allow to derive dynamical dissipative equations of motion for field expectation values $\Phi(x)$. We will start to investigate the analytic structure of $\Gamma[\Phi]$ in the sector where it is best understood, namely the sector of two-point correlation functions.

\subsection{Analytic continuation of two-point functions}
\label{sec:ACTwoPoint}
We discuss now the analytic structure of connected two-point correlation functions which correspond to the second functional derivatives of the Schwinger functional $W[J]$ and the effective action $\Gamma[\Phi]$. For this subsection we consider global equilibrium states with temperature $T$ and fluid velocity $u^\mu$ in flat space with Minkowski metric $g_{\mu\nu} = \eta_{\mu\nu}$. The description is simplest in the coordinate system where the fluid is at rest, i.e. $u^\mu=(1,0,0,0)$ and we will adopt this choice for the time being. Some results can later be taken over to more general local equilibrium states and curved space within the corresponding approximations.

The analytic structure of two-point correlation functions in global equilibrium states is rather well understood, see for example \cite{Abrikosov}. A comprehensive compilation of the main classical results such as the fluctuation-dissipation relation, Onsagers relations and the spectral representation including their derivation is provided in appendix \ref{sec:appA}. Here we discuss the main elements needed for analytic continuation.

From the Schwinger functional in eq.\ \eqref{eq:defSchwingerFunct} one can obtain the two-point function in the Euclidean domain as
\begin{equation}
W_E^{(2)}(x,y) = \frac{\delta^2}{\delta J_a(x) \delta J_b(y)} W_E[J] = \langle \phi_a(x) \phi_b(y) \rangle_c = \langle \phi_a(x) \phi_b(y) \rangle - \langle \phi_a(x) \rangle \langle \phi_b(y) \rangle .
\label{eq:W201}
\end{equation}
The arguments $x$ and $y$ consist here of (real) spatial positions and imaginary times on the Matsubara torus. We write this as $x=(-i\bar x^0, \vec x)$ with $\bar x^0 \in[0,1/T]$ etc. When evaluated for vanishing source $J$ and because of translational symmetry in space and imaginary time, the correlation function in \eqref{eq:W201} is a function of $x-y$ only,
\begin{equation}
\frac{\delta^2}{\delta J_a(x) \delta J_b(y)} W_E[J] {\big |}_{J=0}= \Delta^M_{ab}(x-y).
\label{eq:AC03}
\end{equation}
It is useful to consider the Fourier representation defined by
\begin{equation}
\Delta^M_{ab}(x-y) = T \sum_{n=-\infty}^\infty \int \frac{d^{d-1}p}{(2\pi)^{d-1}} e^{-i \omega_n (\bar x^0-\bar y^0) +i \vec p (\vec x - \vec y)} \, \Delta^M_{ab}\left(i\omega_n, \vec p \right) ,
\end{equation}
where $\omega_n$ is the Matsubara frequency with values $\omega_n=2\pi T n$ for bosonic fields $\phi_a$, $\phi_b$ and $\omega_n=2\pi T (n+1/2)$ for fermionic fields.

As discussed in appendix \ref{sec:appA}, the Matsubara correlation function $\Delta^M_{ab}$ is actually a special case of a more general correlation function $G_{ab}$ which is defined for general complex frequency argument. More specific, on the axis of imaginary Matsubara frequencies one has
\begin{equation}
\Delta^M_{ab}(i\omega_n, \vec p \, ) = G_{ab}(i\omega_n, \vec p \, ),
\end{equation}
and $G_{ab}(p^0, \vec p \, )$ is the unique analytic continuation of $\Delta^M_{ab}(i\omega_n, \vec p \, )$ to the full plane of complex frequencies $p^0 \in \mathbb{C}$ which is analytic everywhere except for possible poles and brach cuts on the real frequency axis $p^0\in \mathbb{R}$. The complex argument Greens function has a K\"{a}llen-Lehmann spectral representation which makes its analytic structure with respect to $p^0$ directly apparent,
\begin{equation}
G_{ab}\left(p^0, \vec p \right) = \int_{- \infty}^\infty dw \;  \frac{\rho_{ab}(w^2- \vec p^2, w)}{w-p^0}.
\end{equation}
see also eq.\ \eqref{eq:GSpectralRep}.

Other correlations functions such as retarded, advanced or Feynman propagator functions follow from the complex frequency Greens function by evaluating it in particular regions in the complex frequency plane, 
\begin{equation}
\begin{split}
\Delta^R_{ab}(p) = & G_{ab}\left(p^0 + i \epsilon, \vec p \,\right), \\
\Delta^A_{ab}(p) = & G_{ab}\left(p^0 - i \epsilon, \vec p \,\right), \\
\Delta^F_{ab}(p) = & G_{ab}\left(p^0 + i \epsilon\;  \text{sign}\left(p^0\right), \vec p \,\right),
\end{split}
\label{eq:RAFfromG}
\end{equation}
see also eq.\ \eqref{eq:RetAdvFeynMatFromCompArgGreensFunc}. These relations show that the complex argument Greens function is actually a very useful object. 

One can actually directly extend the definition of the Schwinger functional $W_E[J]$ by analytic continuation in such a way that the functional derivatives yield the complex argument Green's function \cite{Floerchinger:2011sc}. In momentum space,
\begin{equation}
\frac{\delta^2}{\delta J_a(-p) \delta J_b(q)} W_E[J]{\big |}_{J=0} =G_{ab}(p) \; (2\pi)^4 \delta^{(4)}(p-q).
\label{eq:AC05}
\end{equation}
Obviously, when evaluated in the appropriate regions of the complex frequency plane, this definition agrees with \eqref{eq:AC03} but it has the advantage that one can also directly obtain other correlation functions, e. g. the retarded one, from the general expression \eqref{eq:AC05}.

So far we have worked in the reference frame where the fluid is at rest, $u^\mu=(1,0,0,0)$. Obviously, objects like the complex argument Green's function $G_{ab}(p)$ can also be evaluated in other frames. In that case one should take the combination $\omega=-u^\mu p_\mu$ to be complex in general, and the poles and brach cuts will be on the axis where $\omega$ is real. The momentum components orthogonal to the fluid velocity $\Delta^\mu_{\;\;\nu}p^\nu = (\delta^\mu_{\;\;\nu} + u^\mu u_\nu) p^\nu$ are real. In the following we will work in such a more general frame.

Let us now turn to the effective action $\Gamma_E[\phi]$ as defined in eq.\ \eqref{eq:defGamma1PI} and in particular its second functional derivative. From the definition as a Legendre transform in the Euclidean domain and assuming that $W[J]$ is strictly convex, it follows that
\begin{equation}
\sum_b\int_y \left( \frac{\delta^2 W_E[J]}{\delta J_a(x) \delta J_b(y)} \right) \left( \frac{\delta^2 \Gamma_E[\Phi]}{\delta \Phi_b(y) \delta \Phi_c(z)} \right) = \delta_{ac} \, \delta(x-z) .
\label{eq:Gamma2W2}
\end{equation}
The expectation value $\Phi$ is related to the source $J$ by the field equation \eqref{eq:GammaFieldEquation}.
When evaluated for homogeneous field $\Phi$, one can write the second functional derivative of $\Gamma_E[\Phi]$ in momentum space as
\begin{equation}
\frac{\delta^2}{\delta \Phi_a(-p) \delta \Phi_b(q)} \Gamma_E[\Phi] =P_{ab}(p) \; (2\pi)^4 \delta^{(4)}(p-q).
\label{eq:EALP02}
\end{equation}
A priori this is only defined in the Euclidean domain, i.\ e.\ for imaginary Matsubara frequencies $\omega=-u^\mu p_\mu = i \omega_n$ (as well as real momenta $\Delta^\mu_{\;\;\nu}p^\nu$ orthogonal to the fluid velocity). However, similar to \eqref{eq:AC05} one can extend this to general complex frequencies $\omega$ by analytic continuation.

Due to the relation \eqref{eq:Gamma2W2}, the object $P_{ab}(p)$ satisfies
\begin{equation}
\sum_ b G_{ab}(p) P_{bc}(p) = \delta_{ac} ,
\end{equation}
and is therefore called the inverse complex argument Green's function.
Because of the analytic structure of $G_{ab}(p)$, the eigenvalues of $P_{ab}(p)$ cannot have any zero crossings or branch cuts except on the axis of real $\omega = - u^\mu p_\mu$. One can decompose the inverse complex-argument two-point function
\begin{equation}
P_{ab}(p) = P_{1,ab}(p) - i s_\text{I}(-u^\mu p_\mu) \, P_{2,ab}(p) ,
\label{eq:DecomposeInverseComplexArgGFtext}
\end{equation}
where $s_\text{I}(\omega) = \text{sign}(\text{Im} \; \omega)$. Both functions $P_{1,ab}(p)$ and $P_{2,ab}(p)$ are regular when crossing the real frequency axis. However, the sign $s_\text{I}(-u^\mu p_\mu)$ changes, which leads to a branch cut behavior for the function $P_{ab}(p)$. The function $P_{2,ab}(p)$ parametrizes the strength of the branch cut. From the definition \eqref{eq:EALP02} one obtains the symmetry properties (assuming bosonic fields)
\begin{equation}
P_{ab}(p) = P_{ba}(-p), \quad\quad P_{1,ab}(p) = P_{1,ba}(-p), \quad\quad P_{2,ab}(p) = - P_{2,ba}(-p).
\end{equation}

So far, the analytic continuation of correlation functions was done in momentum and frequency space by analytic continuation from the discrete, imaginary Matsubara frequencies to the plane of complex frequencies $p^0\in \mathbbm{C}$. One would like to have also a position space representation which is not only defined in the configuration space with compact imaginary time direction but can be directly evaluated in real space. To that end one can define the real space representations of the functions $P_1$ and $P_2$ by
\begin{equation}
P_{1,ab}(x-y) = \int \frac{d^dp}{(2\pi)^d} e^{ip(x-y)} P_{1,ab}(p), \quad\quad\quad  P_{2,ab}(x-y) = -i \int \frac{d^dp}{(2\pi)^d}e^{ip(x-y)} P_{2,ab}(p).
\end{equation}
where the momentum and frequency integrals go along the real directions. No ambiguities arise here because $P_{1,ab}(p)$ and $P_{2,ab}(p)$ are regular at the real frequency axis. Note that we have included a factor $-i$ in the definition of $P_{2,ab}(x-y)$ for convenience.
In real position space one can now write the decomposition of the inverse Green's function 
\eqref{eq:DecomposeInverseComplexArgGFtext} as
\begin{equation}
P_{ab}(x-y) = P_{1,ab}(x-y) + s_\text{R}\left(u^\mu \tfrac{\partial}{\partial x^\mu} \right) P_{2,ab}(x-y).
\end{equation}
We have also replaced here
\begin{equation}
s_\text{I}\left(-u^\mu p_\mu \right) = \text{sign}\left(\text{Im}(-u^\mu p_\mu) \right) \to \text{sign}\left(\text{Im}\left(iu^\mu \tfrac{\partial}{\partial x^\mu}\right) \right) = \text{sign}\left(\text{Re}\left(u^\mu \tfrac{\partial}{\partial x^\mu}\right) \right)= s_\text{R}\left(u^\mu \tfrac{\partial}{\partial x^\mu}\right) .
\end{equation}
Typically $P_{1,ab}(p)$ and $P_{2,ab}(p)$ will be a polynomial in combinations of momenta $p^\mu$ and accordingly the position space representations $P_{1,ab}(x-y)$ and $P_{2,ab}(x-y)$ will consist of combinations of distributions such as $\delta(x-y)$ and derivatives thereof. 

We define now also an analytically continued effective action in real position space via a deformation of the time integration contour. More specific, in the Euclidean domain, the time was integrated along the (compact) imaginary direction $dx^0=-id\bar x^0$, see the discussion around \eqref{eq:dSigmamudxmu}. This contour can be deformed such that it goes along the (non-compact) real direction $dx^0\in \mathbb{R}$. The infinitesimal volume element $d^dx \sqrt{g}$ is then real, as well. It is convenient to divide by the factor $i$ that has been introduced in \eqref{eq:complextimeJphi} and to define the {\it analytically continued action in real space} as
\begin{equation}
\Gamma[\Phi] = -i \Gamma_E[\Phi]{\Big |}_\text{deformed time contour}.
\end{equation}
For the effective action in real space we will use standard functional derivatives which differ formally from the functional derivatives in the Euclidean domain by a factor $i$. However, this should not lead to any confusion is practice. 

As an example, the part of the effective action $\Gamma[\Phi]$ that is quadratic in the fields $\delta \Phi$ is of the form
\begin{equation}
\begin{split}
\Gamma_{2} = & \frac{1}{2} \int_{x,y} \delta\Phi_a(x) \left[ P_{1,ab}(x-y) + s_\text{R}\left( u^\mu\tfrac{\partial}{\partial x^\mu}\right)  P_{2,ab}(x-y) \right] \delta \Phi_b(y),\\
= & \frac{1}{2} \int_{x,y} \delta\Phi_a(x) \left[ P_{1,ab}(x-y) + P_{2,ab}(x-y) s_\text{R}\left( u^\mu\tfrac{\partial}{\partial y^\mu}\right) \right] \delta \Phi_b(y),
\end{split}
\label{eq:Gamma2}
\end{equation}
where the integrals over $x$ and $y$ are now in real coordinate space. While the first term $\sim P_1$ in \eqref{eq:Gamma2} is of the standard form, the second term $\sim P_2$ is less common. It parametrizes the discontinuity of the inverse propagator along the real frequency axis. 

If one considers $s_\text{R}(\cdot)$ to be an odd function of the argument, one can actually perform partial integration to transfer the operator $s_\text{R}(u^\mu \partial_\mu)$ between the different terms. This has been done in the second line of \eqref{eq:Gamma2}. In the present context, partial integration is possible because we work in cartesian coordinates and because $u^\mu$ is constant. In a more general coordinate system with differential volume element $d^d x \sqrt{g}$ one can see that partial integration according to 
\begin{equation}
\int d^dx \sqrt{g} \, A(x) [u^\mu(x) \partial_\mu]^N B(x) = (-1)^N \int d^dx \sqrt{g} \, \left\{ [u^\mu(x) \partial_\mu]^N A(x) \right\}  B(x)
\end{equation}
is possible for $\nabla_\mu u^\mu(x)=0$. This could imply that relations like 
\begin{equation}
\int d^dx \sqrt{g} \, A(x) s_\text{R}(u^\mu(x) \partial_\mu) B(x) = - \int d^dx \sqrt{g} \, \left\{ s_\text{R}(u^\mu(x) \partial_\mu) A(x) \right\}  B(x)
\label{eq:partialIntegrationSR}
\end{equation}
could only be used for $\nabla_\mu u^\mu=0$. However, the sign function is unchanged if the argument is rescaled by a positive factor $e^\alpha$ and one can therefore replace $s_\text{R}(u^\mu \partial_\mu)$ by $s_\text{R}(e^\alpha u^\mu \partial_\mu)$ with $\alpha(x)$ chosen such that $\nabla_\mu(e^\alpha u^\mu) = 0$. This shows that $s_\text{R}(u^\mu\partial_\mu)$ is in this sense well defined also in such more general situations.

A more clear geometric picture is obtained if one replaces 
\begin{equation}
s_\text{R}(u^\mu \partial_\mu) \to s_\text{R}({\cal L}_u)
\end{equation}
or $s_\text{R}(u^\mu \partial_\mu) \to s_\text{R}({\cal L}_\beta)$ where ${\cal L}_u$ and ${\cal L}_\beta$ are Lie derivatives in the direction $u^\mu$ and $\beta^\mu=u^\mu/T$, respectively. When acting on scalar fields one has ${\cal L}_u = u^\mu \partial_\mu$ but the Lie derivative is also well defined for vector, tensor or spinor  valued expressions. Moreover, it does not depend on the metric connection. Working with ${\cal L}_\beta$ might sometimes have advantages for calculating variations of fields with fixed $\beta^\mu(x)$.

Note that according to \eqref{eq:RAFfromG} and because of \eqref{eq:Gamma2W2} one can obtain the operators inverse to the retarded and advanced Green's functions in position space from \eqref{eq:Gamma2} by specific choices of the sign $s_\text{R}(u^\mu\partial_\mu)$. One has
\begin{equation}
\begin{split}
\sum_b\int_y \left[ P_{1,ab}(x-y) + P_{2,ab}(x-y) \right] \Delta^R_{bc}(y-z) = & \, \delta_{ac} \delta(x-z) , \\
\sum_b\int_y \left[ P_{1,ab}(x-y) - P_{2,ab}(x-y) \right] \Delta^A_{bc}(y-z) = & \, \delta_{ac} \delta(x-z) ,
\end{split}
\label{eq:RAInverse}
\end{equation}
so that these the two combinations of $P_1$ and $P_2$ on the left hand side of \eqref{eq:RAInverse} can be understood as the inverse retarded and advanced propagators, respectively.

Let us now come to the implications of the field equation \eqref{eq:GammaFieldEquation} after analytic continuation. To study its implications we expand the effective action $\Gamma_E[\phi]$ in powers of fields
\begin{equation}
\Gamma[\Phi] = \int_p \frac{1}{2} \Phi_a(-p) P_{ab}(p) \, \Phi_b(p) + \Delta\Gamma[\Phi],
\label{eq:EAL05}
\end{equation}
where $\Delta\Gamma[\Phi]$ contains terms of cubic and higher order in $\Phi$. 
We assume here that expectation value $\Phi$ vanishes for vanishing source which implies that no linear term is present in eq.\ \eqref{eq:EAL05}. We have also dropped a possible constant term as irrelevant for the following discussion. The field equation \eqref{eq:GammaFieldEquation} reads now formally
\begin{equation}
P_{ab}(p) \Phi_b(p) + \frac{\delta }{\delta \Phi_a(-p)} \Delta\Gamma[\Phi] = 0 .
\label{eq:FEAnalyCont}
\end{equation}
To define the objects in \eqref{eq:FEAnalyCont}, we have used here analytic continuation from the Euclidean domain to the Minkowski signature domain, such that the frequency $p^0$ is real. However, this leaves open what has to be taken for $s_\text{I}(-u^\mu p_\mu)$. From causality arguments, the linear part of the field equation for forward time evolution should be inverse to the retarded propagator (and for backward time evolution inverse to the advanced propagator). This suggests that one has to take for forward time evolution variations $\delta\Phi_a(-p)$ with $s_\text{I}(-u^\nu p_\nu)=+1$ and for backward time evolution $s_\text{I}(-u^\nu p_\nu)=-1$. This agrees with the generalized variational principle introduced in sect.\ \ref{sec:DampedHO}. The result is
\begin{equation}
\begin{split}
\left(P_{1,ab}(p) -i P_{2,ab}(p)\right)\Phi_b(p) + \frac{\delta }{\delta \Phi_a(-p)} \Delta\Gamma[\Phi] = 0 & \quad\quad\quad \text{(forward time evolution)}, \\
\left(P_{1,ab}(p) + i P_{2,ab}(p) \right) \Phi_b(p) + \frac{\delta }{\delta \Phi_a(-p)} \Delta\Gamma[\Phi] = 0 & \quad\quad\quad \text{(backward time evolution)}.
\end{split}
\label{eq:EOMFieldTheory}
\end{equation}
It is straight forward to translate these relations to position space. The dissipative equations of motion will be discussed in more general context also in section \ref{sec:GeneralizedVariationalPrinciple}.

\subsection{Analytic structure of higher order correlation functions}
\label{sec:ACHigherOrderCF}

So far we have studied the analytic structure of two-point functions or the corresponding sector of the effective action $\Gamma[\Phi]$. In general, the analytic structure of higher order correlation functions is much less understood. 

It is, however, nevertheless possible to generalize the method to somewhat more general situations. Consider an effective action of the following form (we take here $\phi$ and $\chi$ to be bosonic fields for simplicity)
\begin{equation}
\begin{split}
\Gamma[\phi,\chi] = & \int_p \frac{1}{2} \phi_a(-p) P_{ab}(p) \phi_b(p) + \int_p \frac{1}{2} \chi_a(-p) Q_{ab}(p) \chi_b(p)\\
& + \int_{q_1,q_2,q_3}  \delta(q_1+q_2+q_3) \; \frac{1}{2} h_{abc}(q_2,q_3) \chi_a(q_1) \phi_b(q_2) \phi_c(q_3) + \Delta \Gamma[\phi] ,
\end{split}
\label{eq:GammaEx}
\end{equation}
where $h_{abc}(q_2,q_3) = h_{acb}(q_3,q_2) $. The matrix $P_{ab}(p) = P_{ba}(-p)$ can be decomposed as in eq.\ \eqref{eq:DecomposeInverseComplexArgGFtext} and similarly $Q_{ab}(p) = Q_{ba}(-p)$. The effective vertex $h_{abc}(q_2,q_3)$ and the terms in $\Delta \Gamma[\phi]$ are assumed not to contain any discontinuous terms involving $s_\text{I}(-u^\mu p_\mu)$ (where $p$ is some combination of the involved momenta). The equations of motion following from \eqref{eq:GammaEx} for forward (backward) time evolution are obtained by following the general principle described above, as
\begin{equation}
\begin{split}
& \left( P_{1,ab}(p) \mp i P_{2,ab}(p) \right) \phi_b(p) \\
& + \int\limits_{q_1, q_3} \delta(q_1-p+q_3) \; h_{dab}(-p,q_3) \chi_d(q_1) \phi_b(q_3) + \frac{\delta}{\delta\phi_a(-p)}\Delta\Gamma[\phi] = 0,
\end{split}
\label{eq:GammaExEOM1}
\end{equation}
as well as
\begin{equation}
\begin{split}
& \left( Q_{1,ab}(p) \mp i Q_{2,ab}(p) \right) \chi_b(p)  + \int\limits_{q_2, q_3} \delta(-p+q_2+q_3) \; \frac{1}{2} h_{abc}(q_2,q_3) \phi_a(q_2) \phi_b(q_3) = 0. 
\end{split}
\label{eq:GammaExEOM2}
\end{equation}
Because equation \eqref{eq:GammaExEOM2} is linear in $\chi$ it can formally be solved,
\begin{equation}
\chi_a(p) = - \left( Q_{1}(p) \mp i Q_{2}(p) \right)^{-1}_{ad} \int\limits_{q_2,q_3} \delta(-p+q_2+q_3) \; \frac{1}{2}h_{dbc}(q_2,q_3) \phi_b(q_2) \phi_c(q_3) .
\end{equation}
This solution, in turn, can be used in eq.\ \eqref{eq:GammaExEOM1} to give
\begin{equation}
\begin{split}
& \left( P_{1,ab}(p) \mp i P_{2,ab}(p) \right) \phi_b(p) \\
& - \int\limits_{q_2, q_3, q_4} \delta(-p+q_1+q_2+q_3) \; h_{eab}(-p,q_2) \left( Q_{1}(p) \mp i Q_{2}(p) \right)^{-1}_{ef}
h_{fcd}(q_3,q_4) \; \phi_b(q_2) \phi_c(q_3) \phi_d(q_4) \\
& + \frac{\delta}{\delta\phi_a(-p)}\Delta\Gamma[\phi] = 0.
\end{split}
\label{eq:GammaExEOM3}
\end{equation}
Interestingly, in this form, the equation of motion for $\phi$ has now a non-linear term with a discontinuity. The field equation is such that the matrix $\left( Q_{1}(p) \mp i Q_{2}(p) \right)^{-1}_{ef}$ which connects $\phi_c(q_3)\phi_d(q_4)$ to $\phi_b(q_2)$ and the point of variation $\delta \phi_a(-p)$ is the retarded (advanced) propagator for forward (backward) time evolution. That is what one expects from causality considerations.

It is now illuminating to note that the field $\chi$ enters quadratic in the effective action $\Gamma[\phi, \chi]$ in eq.\ \eqref{eq:GammaEx} and one can therefore integrate it out directly. That gives
\begin{equation}
\begin{split}
\Gamma[\phi] & =  \int_p \frac{1}{2} \phi_a(-p) P_{ab}(p) \phi_b(p) \\
&  - \int\limits_{q_1\ldots q_4} \delta(q_1+q_2+q_3+q_4)  \frac{1}{8} h_{eab}(q_1,q_2)  \left( Q(q_3+q_4) \right)^{-1}_{ef} h_{fcd}(q_3,q_4) \; \phi_a(q_1) \phi_b(q_2) \phi_c(q_3) \phi_d(q_4) \\
& + \Delta \Gamma[\phi] .
\end{split}
\end{equation} 
Note that the second term is quartic in $\phi$ and contains the symbol $s_\text{I}(\omega_3+\omega_4) = - s_\text{I}(\omega_1+\omega_2)$ where $\omega_1 = - u_\nu p^\nu_1$ etc. In more general situations, the effective action $\Gamma[\Phi]$ might contain terms involving $s_\text{I}(\omega)$ where $\omega$ is any combination of frequencies of fields.

\subsection{The general structure of the analytic quantum effective action}
\label{sec:GeneralStructureAnalyticEffectiveAction}

Let us now discuss the expected structure of the analytically continued quantum effective action in more detail. For convenience, we will use a general coordinate system and one requirement for the terms that appear in $\Gamma[\Phi]$ will be general covariance (we neglect the effect of possible gravitational anomalies in this exploratory study). Obviously, $\Gamma[\Phi]$ can have contributions $\Gamma_\text{Regular}[\Phi]$ which do not show any discontinuities along the real frequency axis. Such terms can be treated in the standard way, very similar to Euclidean field theory where discontinuities do not arise at all. 

However, in addition there can be terms that are discontinuous along the real frequency axis. We will assume that such terms are written in terms of the symbol $s_\text{I}(\omega)$ in frequency representation or involve the symbol 
\begin{equation}
s_\text{R}\left(u^\mu(x) \tfrac{\partial}{\partial x^\mu} \right) \qquad \qquad \text{or} \qquad \qquad s_\text{R}\left({\cal L}_u \right)
\end{equation}
in a position space representation.\footnote{It might be a bit surprising to see the fluid velocity $u^\mu$ appearing, in particular if one is actually interested in starting with the conventional vacuum state $T\to 0$. For this state, the fluid velocity is not well defined. We expect that it does for this state actually not play a role in which direction $u^\mu$ is pointing initially, as long as it is time-like and future oriented. Within a local equilibrium assumption, a fluid with non-zero temperature and well defined fluid velocity is formed, however, as soon as some energy has been dissipated.} The position space representation has the advantage that it can be written in an explicitly covariant way more easily. 

The discontinuous terms that have been discussed above were of the form
\begin{equation}
\Gamma_\text{Disc}[\Phi] = \int d^d x \sqrt{g} \; \left\{ f[\Phi](x) \;  s_\text{R}{\big (}u^\mu(x) \tfrac{\partial}{\partial x^\mu}{\big )} \; g[\Phi](x) \right\}
\label{eq:DiscStruct2}
\end{equation}
The objects $f[\Phi](x)$ and $g[\Phi](x)$ can for example be local functionals linear in $\Phi(x)$ as in the example of the damped harmonic oscillator discussed in section \ref{sec:DampedHO} or more general linear functionals of the fields as discussed in section \ref{sec:ACTwoPoint}. However, they can also be non-linear functionals in $\Phi$ as discussed in section \ref{sec:ACHigherOrderCF}. In the situations discussed so far, $f[\Phi](x)$ and $g[\Phi](x)$ were regular in the sense that they were assumed not to involve further discontinuities. Note that one can transfer the operator $s_\text{R}{\big (}u^\mu(x) \tfrac{\partial}{\partial x^\mu}{\big )}$ from acting on $g[\Phi](x)$ to acting on $f[\Phi](x)$ by partial integration, see \eqref{eq:partialIntegrationSR} and the discussion there.

Other structures that are conceivable are for example of the form (we use the abbreviation $\int_x = \int d^d x \sqrt{g}$ etc.)
\begin{equation}
\Gamma_\text{Disc}[\Phi] = \int_{x,y}  \; \left\{ f[\Phi](x) \;  s_\text{R}{\big (}u^\mu(x) \tfrac{\partial}{\partial x^\mu} {\big )} \; g[\Phi](x,y) \;  s_\text{R}{\big (} u^\mu(y) \tfrac{\partial}{\partial y^\mu} {\big )} \; h[\Phi](y)  \right\},
\end{equation}
or 
\begin{equation}
\begin{split}
\Gamma_\text{Disc}[\Phi] = \int_{x,y,z}  \; {\Big \{} & f[\Phi](x) \;  s_\text{R}{\big (}u^\mu(x) \tfrac{\partial}{\partial x^\mu} {\big )} \; g[\Phi](x,y,z) \;  s_\text{R}{\big (} u^\mu(y) \tfrac{\partial}{\partial y^\mu} {\big )} \; h[\Phi](y) \\
 & + s_\text{R}{\big (} u^\mu(z) \tfrac{\partial}{\partial z^\mu} {\big )} \;  j[\Phi](z) 
{\Big \}},
\end{split}
\label{eq:treeStrctDisc}
\end{equation}
where $f[\Phi]$, $g[\Phi]$, $h[\Phi]$ and $j[\Phi]$ are linear or non-linear functionals of the field expectation value $\Phi$ without discontinuities. We assume here that they are scalars in the sense of general covariance but it might also be possible to allow for vectors, spinors or tensors; possibly by using covariant or Lie derivatives within the operator $s_\text{R}(\ldots)$. The discussion of the generalized variational principle in the next section below will allow for general ``tree structures'' as in the examples above, with one spatial integral and a corresponding discontinuity operator $s_\text{R}$ for each line and functionals of the field $\Phi$ without discontinuities at the nodes. However, this structure should not contain any ``loops''.

It has to be stated that at this point it is merely a conjecture that possible terms in the analytically continued quantum effective action are all of the form described above. More detailed investigations are needed to confirm this or to find more general allowed structures. In practice, it is typically not possible to trace the full form of the effective action and the most important terms involving discontinuities arise at quadratic order in the fields (as discussed in section \ref{sec:ACTwoPoint}) or are of the form \eqref{eq:DiscStruct2}. 

\subsection{Relation to time-ordered and Matsubara actions}
\label{sec:RelationToTimeOrderedMatsubara}

Here we discuss how the analytically continued effective action is related to other effective actions. First, the time-ordered or Feynman propagator follows from the complex frequency propagator by choosing the appropriate integration contour in the complex frequency plane, cf.\ eq.\ \eqref{eq:RAFfromG}. This amounts to replacing $s_\text{I}(\omega) \to s_\text{R}(\omega)$. The same replacement yields the time-ordered inverse propagator from the complex frequency inverse propagator. Moreover, the discussion in the previous subsections shows that this prescription can actually be used for the entire action. 

In time representation the replacing $s_\text{I}(\omega) \to s_\text{R}(\omega)$ becomes formally $s_\text{R}(\partial_t) \to - s_\text{I}(\partial_t)$. (We assume $u^\mu=(1,0,0,0)$ in this subsection for simplicity.) However, this should rather be understood as a non-local operator in the following sense. In frequency domain $s_\text{R}(\omega)$ corresponds to the operator (seen as a matrix with indices $\omega$ and $\omega^\prime$)
\begin{equation}
O(\omega,\omega^\prime) = 2\pi \delta(\omega^\prime - \omega) \text{sign}(\omega)
\end{equation}
In the time domain the corresponding operator is
\begin{equation}
\begin{split}
O(t^\prime, t) = & \int \frac{d\omega^\prime}{2\pi} e^{-i\omega^\prime t^\prime} \int \frac{d\omega}{2\pi} e^{i\omega t} O(\omega^\prime, \omega) = \int \frac{d \omega}{2\pi} e^{-i\omega(t^\prime- t)} \; \text{sign}(\omega) \\
= & - \frac{i}{\pi} \mathsf{PV} \frac{1}{t^\prime-t} = - \delta(t^\prime-t) s_\text{I}(\partial_t).
\end{split}
\label{eq:nonlocalsI}
\end{equation}
In the second to last equation $\mathsf{PV}$ denotes the principal value and the last equation defines more formally what is meant by the symbol $s_\text{I}(\partial_t)=\text{sign}(\text{Im} \, \partial_t)$. Keeping this correspondence in mind we will continue to work with the symbolic notation $s_\text{I}(\partial_t)$ in the following.

Symbolically, in terms of the effective Lagrangian density (defined by $\Gamma[\phi] = - \int d^dx \sqrt{g} \, \mathscr{L}$), the Feynman effective action is obtained from the analytic effective action by
\begin{equation}
\mathscr{L}_F = \mathscr{L} {\big |}_{s_\text{R}(\partial_t) \to - s_\text{I}(\partial_t)} .
\end{equation}
In a completely analogous way one obtains the anti-time-ordered or Dyson effective action by
\begin{equation}
\mathscr{L}_D = \mathscr{L} {\big |}_{s_\text{R}(\partial_t) \to s_\text{I}(\partial_t)} .
\end{equation}

The Euclidean or Matsubara effective action is obtained by analytic continuation to Euclidean frequencies and replacing $s_\text{I}(\omega) \to s_\text{I}(i \omega_n) = s_\text{R}(\omega_n)$, where $\omega_n$ is the Masubara frequency, 
\begin{equation}
\mathscr{L}_M = \mathscr{L} {\big |}_{s_\text{I}(\omega) \to s_\text{R}(\omega_n)} .
\end{equation}
In (Euclidean) time representation this corresponds formally to $s_\text{R}(\partial_t)\to -s_\text{I}(\partial_\tau)$. However, in practical applications of the imaginary time formalism such as for tunneling problems, one may rather work with a non-local representation analogous to \eqref{eq:nonlocalsI}, see ref.\ \cite{Caldeira:1982uj} for an example.

In this context, it is actually useful to know the generalization of \eqref{eq:nonlocalsI} for a compact time direction where $\tau=\tau+1$,
\begin{equation}
\begin{split}
\sigma(\tau^\prime-\tau) = & \sum_{n\neq0} e^{-i2\pi n ( \tau^\prime- \tau)} \; \text{sign}(n)
= \frac{e^{2\pi i(\tau^\prime-\tau)}-e^{-2\pi i(\tau^\prime-\tau)}}{e^{2\pi i(\tau^\prime-\tau)}+e^{-2\pi i(\tau^\prime-\tau)}-2}.
\end{split}
\label{eq:nonlocalsIcompact}
\end{equation}
Note in particular that $\sigma(\tau^\prime-\tau) = - \sigma(\tau-\tau^\prime)$ is anti-symmetric.
 
The Schwinger-Keldysh or closed-time-path effective action can be constructed via the difference of time-ordered and anti-time-ordered actions or directly from advanced and retarded inverse propagators. A more detailed discussion will be given elsewhere.

\section{Generalized variational principle and dissipative equations of motion}
\label{sec:GeneralizedVariationalPrinciple}

We now formulate the variational principle by which one can obtain the dissipative, causal equations of motion for the field expectation value $\Phi_a(x)$ from the analytically continued quantum effective action $\Gamma[\Phi]$. The main question here is how the sign $s_\text{R}(u^\mu\partial_\mu)$ has to be chosen.

We formulate the variational principle such that it leads to the equations of motion for (standard) forward time propagation. The equations for a (somewhat artificial) situation where the time direction is reversed can be obtained with the appropriate changes of sign in a straight forward way. For an effective action as described in subsection \eqref{sec:GeneralStructureAnalyticEffectiveAction} one can obtain the equation of motion for the field $\Phi_a(x)$ from the equation
\begin{equation}
\frac{\delta \Gamma[\Phi]}{\delta \Phi_a(x)} {\bigg |}_\text{ret} = \sqrt{g}(x) \; J_a(x) .
\label{eq:FieldEqRet}
\end{equation}
To take the ``retarded variational derivative'' one calculates the variation $\delta \Gamma[\Phi]$ as usual and sets then $s_\text{R}(u^\mu \tfrac{\partial}{\partial x^\mu}) \to -1$ if the field variation $\delta \Phi_a(x)$ is to the right of that operator and $s_\text{R}(u^\mu \tfrac{\partial}{\partial x^\mu}) \to 1$ if the field variation $\delta \Phi_a(x)$ is to the left of that operator. This principle is consistent with the possibility to transfer $s_\text{R}(u^\mu \tfrac{\partial}{\partial x^\mu})$ by partial integration, see the discussion around eq.\ \eqref{eq:partialIntegrationSR}.

For tree-like structures like in \eqref{eq:treeStrctDisc} one has to choose $s_\text{R}(u^\mu \partial_\mu) \to -1$ if the derivative operator points towards the node $f[\Phi](x)$, $g[\Phi](x,y,z)$ or similar that is varied, and $s_\text{R}(u^\mu \partial_\mu) \to 1$ if the derivative operator acts to the opposite direction. For tree-like structures this is well defined because every node is connected to all other nodes in a uniquely orientable way. As examples for the prescribed variational principle one may take the harmonic oscillator in section \ref{sec:DampedHO} or the example with composite fields in subsection \ref{sec:ACHigherOrderCF}.

A particularly simple situation arises when the symbol $s_\text{R}(u^\mu\partial_\mu)$ appears within a commutator such as
\begin{equation}
I[\Phi] = \int d^dx \sqrt{g} \, \frac{1}{2}\left[ \Phi_a(x), s_\text{R}(u^\mu\partial_\mu) \right] f_a[\Phi](x),
\end{equation}
where $f[\Phi](x)$ is a regular functional of the fields. In that case, by the rules formulated above, the ``retarded variational derivative'' with respect to $\Phi_a$ hits effectively only the field in the commutator because all other terms cancel. One has
\begin{equation}
\frac{\delta I[\Phi]}{\delta \Phi_a(x)} {\bigg |}_\text{ret} = \sqrt{g}(x)\,  f_a[\Phi](x).
\end{equation}

The ``oriented'' or ``retarded'' functional derivative described above is closely connected with causality. It makes sure that the field equation following from eq.\ \eqref{eq:FieldEqRet} depends only on field expectation values and other physical information in the past light cone with respect to $x$. To see this, take another functional derivative of the field equation \eqref{eq:FieldEqRet} (we specialize to flat space and use here cartesian coordinates with $\sqrt{g}=1$ for simplicity),
\begin{equation}
\frac{\delta}{\delta \Phi_b(y)} \frac{\delta \Gamma}{\delta \Phi_a(x)} {\bigg |}_\text{ret} = \frac{\delta }{\delta \Phi_b(y)} J_a(x) .
\label{eq:42}
\end{equation}
Inverting this equation using eqs.\ \eqref{eq:Gamma2W2} and \eqref{eq:RAFfromG} gives the retarded Green's function,
\begin{equation}
\frac{\delta}{\delta J_b(y)} \Phi_a(x) = \Delta^R_{ab}(x,y).
\label{eq:Causality}
\end{equation}
This equation tells that the expectation value $\Phi_a(x)$ that solves the field equation \eqref{eq:FieldEqRet} is causal in the sense that it can only be modified by sources $J_b(y)$ at positions $y$ that are in the past light cone of $x$. Otherwise, the right hand side of \eqref{eq:Causality} vanishes. Note that the issue of causality is more complicated in curved space-time. One would need more powerful arguments to exclude the appearance of closed time-like curves, for example.

Note that an alternative way to obtain causal and real equations of motion would have been via the Schwinger-Keldysh formalism. The one-particle irreducible effective action can be defined on a closed time path and its variation leads to causal and real equations of motion including dissipative effects \cite{Jordan:1986ug}, see also \cite{Calzetta:1986cq}.  The closed time-path formalism has also been discussed in the context of curved space \cite{Calzetta:1986ey}, including expressions for a covariantly conserved expectation value of the energy-momentum tensor. To that topic we turn next.

\section{Energy-momentum conservation and entropy production}
\label{sec:EnergyMomentumConservationEntropyProduction}

In this section we discuss the important issue of energy and momentum conservation as well as entropy production in the context of an analytically continued quantum field theory with dissipation. We start by deriving and discussing an expression for the expectation value of $T^{\mu\nu}$ from the analytic effective action.

\subsection{Energy-momentum tensor}
\label{sec:EnMomTensor}

Very similar to the field equations, the expectation value of the energy-momentum tensor can contain dissipative terms. In analogy to the retarded derivative \eqref{eq:FieldEqRet} that leads to the equations of motion, the following expression should hold
\begin{equation}
\frac{\delta \Gamma[\Phi, g_{\mu\nu}, \beta^\mu]}{\delta g_{\mu\nu}(x)}{\bigg |}_\text{ret} = - \frac{1}{2} \sqrt{g} \langle T^{\mu\nu}(x) \rangle.
\label{eq:ExpValTmunu0}
\end{equation} 
The ``retarded functional derivative'' is defined as discussed below \eqref{eq:FieldEqRet} but now for variations with respect to $g_{\mu\nu}(x)$. Note that the right hand side of \eqref{eq:ExpValTmunu0} should also appear as the source term for Einsteins field equations in a context where the analytic action $\Gamma[\Phi, g_{\mu\nu}, \beta^\mu]$ contains gravitational terms.  

For the following it is useful to decompose the analytic action like
\begin{equation}
\Gamma[\Phi, g_{\mu\nu}, \beta^\mu] = \Gamma_R[\Phi, g_{\mu\nu}, \beta^\mu] + \Gamma_D[\Phi, g_{\mu\nu}, \beta^\mu] ,
\end{equation}
where the reduced action is defined as an integral over real space
\begin{equation}
\Gamma_R[\Phi,g_{\mu\nu},\beta^\mu] = - \int d^dx \sqrt{g} \, \mathscr{L}_R,
\label{eq:defGammaRed}
\end{equation}
of the effective Lagrangian with non-dissipative terms only,
\begin{equation}
\mathscr{L}_R = \mathscr{L} {\big |}_{s_\text{R}(\partial_t) \to 0} .
\label{eq:defLRed}
\end{equation}
The remainder term $\Gamma_D[\Phi,g_{\mu\nu},\beta^\mu] $ contains dissipative (i.e. discontinuous) terms, only. In a similar way we decompose the expectation value of the energy momentum tensor according to
\begin{equation}
\langle T^{\mu\nu}(x) \rangle = (\bar T_R)^{\mu\nu}(x) + (\bar T_D)^{\mu\nu}(x),
\end{equation}
with
\begin{equation}
\frac{\delta \Gamma_R[\Phi, g_{\mu\nu}, \beta^\mu]}{\delta g_{\mu\nu}(x)} = - \frac{1}{2} \sqrt{g} \,  (\bar T_R)^{\mu\nu}(x), \quad\quad\quad\quad
\frac{\delta \Gamma_D[\Phi, g_{\mu\nu}, \beta^\mu]}{\delta g_{\mu\nu}(x)}{\bigg |}_\text{ret} = - \frac{1}{2} \sqrt{g} \, (\bar T_D)^{\mu\nu}(x).
\label{eq:ExpValTmunuRD}
\end{equation} 
Because $\Gamma_R$ contains no dissipative terms, the retarded functional derivative agrees with the conventional one.

\subsection{General covariance and energy-momentum conservation}
\label{sec:EnergyMomentumConservation}

Let us now discuss energy-momentum conservation. For theories without dissipation, the (covariant) conservation of the energy-momentum tensor is a consequence of general covariance and the field equations. We will discuss this here in terms of the reduced action $\Gamma_R[\Phi, g_{\mu\nu},\beta^\mu]$ as defined in \eqref{eq:defGammaRed} and \eqref{eq:defLRed}.

Infinitesimal general coordinate transformations can be written as a ``gauge transformation'' of the metric (see for example \cite{Weinberg:1972kfs})
\begin{equation}
\delta g^G_{\mu\nu}(x) = g_{\mu\lambda}(x) \frac{\partial \epsilon^\lambda(x)}{\partial x^\nu} +
g_{\nu\lambda}(x) \frac{\partial \epsilon^\lambda(x)}{\partial x^\mu} + \frac{\partial g_{\mu\nu}(x)}{\partial x^\lambda} \epsilon^\lambda(x) ,
\label{eq:GaugeChangeMetric}
\end{equation}
with an infinitesimal vector field $\epsilon^\lambda(x)$. Also the matter field expectation values $\Phi_a(x)$ transform according to their respective representation $\delta \Phi_a^G(x)$. The combined fluid velocity and temperature field $\beta^\mu=u^\mu/T$ transforms as a vector,
\begin{equation}
\delta \beta_G^{\mu}(x) = - \beta^\nu(x) \frac{\partial \epsilon^\mu(x)}{\partial x^\nu} + \frac{\partial \beta^\mu(x)}{\partial x^\nu} \epsilon^\nu(x) .
\label{eq:GaugeChangeTemp}
\end{equation}

The reduced effective action $\Gamma_R[\Phi, g_{\mu\nu},\beta^\mu]$ is invariant under general coordinate transformations (we neglect possible gravitational anomalies), 
\begin{equation}
\Gamma_R[\Phi+\delta\Phi^G,g_{\mu\nu}+\delta g_{\mu\nu}^G,\beta^\mu+\beta^\mu_G] = \Gamma_R[\Phi,g_{\mu\nu},\beta^\mu].
\end{equation}
In a situation without dissipation, the variation of $\Gamma_R$ with respect to the field expectation values $\Phi$ would vanish as a consequence of the equations of motion. If in addition, the dependence on $\beta^\mu(x)$ drops out (for example due to $T=0$), one has
\begin{equation}
\delta \Gamma_R = -\frac{1}{2} \int d^dx \sqrt{g} \, \langle T^{\mu\nu}(x) \rangle \delta g_{\mu\nu}^G(x).
\label{eq:energymomentumreducedaction}
\end{equation}
One can then use eq.\ \eqref{eq:GaugeChangeMetric} to write this as
\begin{equation}
\delta \Gamma_R = \int d^d x \sqrt{g} \, \epsilon^\lambda(x)  \nabla_\mu \langle T^\mu_{\;\;\lambda}(x) \rangle.
\end{equation}
Because $\epsilon^\lambda(x)$ is arbitrary, general covariance implies the covariant conservation law $ \nabla_\mu \langle T^\mu_{\;\;\lambda}(x) \rangle=0$.

In the presence of dissipative effects, the picture is changed at three points. First, the dissipative equations of motion \eqref{eq:FieldEqRet} do not imply that the reduced action $\Gamma_R$ is stationary with respect to variations of $\Phi$. Second, the expectation value of the energy momentum tensor is now given by the retarded functional derivative of the analytic action according to \eqref{eq:ExpValTmunu0} instead of the variation of the reduced action in \eqref{eq:energymomentumreducedaction}. Third, it is now not consistent any more to drop the dependence on $\beta^\mu(x)$. The dissipative effects will generate a heat bath at non-zero temperature even if one starts initially at $T=0$.\footnote{This argument assumes a local equilibrium picture. More general, in a far-from equilibrium situations one will have to take other variables into account that play a role similar to $\beta^\mu$ in the current discussion.} The change of $\Gamma_R$ due to an infinitesimal general coordinate transformation becomes
\begin{equation}
\delta \Gamma_R = \int d^dx \frac{\delta \Gamma_R}{\delta \Phi_a(x)} \delta\Phi_a^G(x) +  \int d^d x \sqrt{g} \, \epsilon^\lambda(x)  \nabla_\mu  (\bar T_R)^\mu_{\;\;\lambda}(x) + \int d^dx \frac{\delta \Gamma_R}{\delta \beta^\mu(x)} \delta\beta^\mu_G(x).
\label{eq:GammaRedGauge}
\end{equation}
In an autonomous situation, the complete energy-momentum tensor $(\bar T_R)^{\mu\nu} + (\bar T_D)^{\mu\nu}$ has to be covariantly conserved so that one can replace
\begin{equation}
\nabla_\mu  (\bar T_R)^\mu_{\;\;\lambda}(x) = - \nabla_\mu  (\bar T_D)^\mu_{\;\;\lambda}(x).
\end{equation}
One is therefore left with
\begin{equation}
\int d^d x \left\{\frac{\delta \Gamma_R}{\delta \Phi_a(x)} \delta\Phi_a^G(x) - \sqrt{g}(x) \, \epsilon^\lambda(x) \nabla_\mu (\bar T_D)^\mu_{\;\;\lambda}(x) + \frac{\delta \Gamma_R}{\delta \beta^\mu(x)} \delta\beta^\mu_G(x) \right\} =0.
\label{eq:515}
\end{equation}
These equations must be fulfilled for arbitrary choice of $\epsilon^\lambda(x)$ which leads to four additional equations. These  differential equations determine the space-time evolution of $\beta^\mu(x)$. Alternatively, one can determine such equations from the covariant conservation law $\nabla_\mu \langle T^{\mu\nu} \rangle=0$ itself.

To analyze the implications of \eqref{eq:515} in a little more detail, define the field $K_\mu(x)$ by
\begin{equation}
 \frac{\delta \Gamma_R}{\delta \beta^\mu(x)} = \sqrt{g}(x) \, K_\mu(x) ,
 \label{eq:defKmu}
\end{equation}
and write, using \eqref{eq:FieldEqRet} for $J_a(x)=0$,
\begin{equation}
\frac{\delta \Gamma_R}{\delta \Phi_a(x)} = - \frac{\delta \Gamma_D}{\delta \Phi_a(x)}{\bigg |}_\text{ret} = - \sqrt{g}(x) \, M_a(x).
\label{eq:defMa}
\end{equation}
For concreteness, assume that $\Phi_a(x)$ transforms as a scalar, 
\begin{equation}
\delta \Phi_a^G(x) = \epsilon^\lambda(x) \frac{\partial}{\partial x^\lambda} \Phi_a(x).
\end{equation}
Equation \eqref{eq:515} becomes then
\begin{equation}
\int d^dx \sqrt{g} \, \epsilon^\lambda(x)  \left\{ - M_a(x) \partial_\lambda \Phi_a(x) - \nabla_\mu (\bar T_D)^\mu_{\;\;\lambda}(x)+   \nabla_\mu \left[\beta^\mu(x) K_\lambda(x) \right]  + K_\mu(x) \nabla_\lambda \beta^\mu(x) \right\} = 0.
\end{equation}
Because $\epsilon^\lambda(x)$ is arbitrary, this implies
\begin{equation}
M_a(x) \partial_\lambda \Phi_a(x) + \nabla_\mu (\bar T_D)^\mu_{\;\;\lambda}(x)=   \nabla_\mu \left[\beta^\mu(x) K_\lambda(x) \right]  + K_\mu(x) \nabla_\lambda \beta^\mu(x).
\label{eq:520}
\end{equation}
As stated above, these four differential equations determine the space-time evolution of $\beta^\mu(x)$ for a physical situation with covariantly conserved energy-momentum tensor.

\subsection{Entropy production}
\label{sec:Entropy}
It is illustrating to analyze the differential equations obtained from general covariance in the previous subsection in a little more detail. In particular, consider the contraction of eq.\ \eqref{eq:520} with $\beta^\lambda$. It can be written as
\begin{equation}
M_a \beta^\lambda \partial_\lambda \Phi_a + \beta^\lambda \nabla_\mu (\bar T_D)^\mu_{\;\;\lambda}= \nabla_\mu \left[ \beta^\mu \beta^\lambda K_\lambda \right] .
\label{eq:ent1}
\end{equation}
To understand the significance of the term on the right hand side of \eqref{eq:ent1}, consider the special case where the only $\beta^\mu$-dependent term in the reduced action $\Gamma_R$ is the effective potential,  
\begin{equation}
 \frac{\delta \Gamma_R}{\delta \beta^\mu(x)} = \frac{\delta}{\delta\beta^\mu(x)} \int d^dx \sqrt{g} \, U(T)
\end{equation}
where $U(T)=-p(T)$ is the grand canonical potential density and $T=1/\sqrt{-g_{\mu\nu}\beta^\mu \beta^\nu}$ is the temperature. In that case $K_\lambda = - s T^3 \beta_\lambda$, where $s=\partial p/ \partial T$ is the entropy density. This implies
\begin{equation}
\beta^\mu \beta^\lambda K_\lambda = s^\mu
\label{eq:defEntropyCurrent}
\end{equation}
where $s^\mu = s u^\mu$ is the entropy current in the present situation. One can actually extend this to a definition of the entropy current $s^\mu$ in more general situations. With this interpretation in mind, equation \eqref{eq:ent1} has very interesting consequences. In general, the divergence $\nabla_\mu s^\mu$ measures the local production of entropy by dissipative processes. The local entropy production by the dissipative processes concerning the scalar field $\Phi_a$ as well as the dissipative terms in the energy-momentum tensor is therefore given by the left hand side of \eqref{eq:ent1}.

Assuming that there are no cancelations between the two terms, one should have as a consequence of a local form of the second law of thermodynamics $\nabla_\mu s^\mu(x) \geq 0$,
\begin{equation}
M_a(x) \beta^\mu(x) \partial_\mu \Phi_a(x) = \frac{1}{\sqrt{g}(x)} \frac{\delta \Gamma_D}{\delta \Phi_a(x)}{\bigg |}_\text{ret} \beta^\mu(x) \partial_\mu \Phi_a(x) \geq 0,
\label{eq:localSecondLaw1}
\end{equation}
and similar
\begin{equation}
\beta_\mu(x) \nabla_\nu (\bar T_D)^{\mu\nu}(x) = \beta_\mu(x) \nabla_\nu \left[ - \frac{2}{\sqrt{g}(x)} \frac{\delta \Gamma_D}{\delta g_{\mu\nu}(x)}{\bigg |}_\text{ret} \right]  \geq 0.
\label{eq:localSecondLaw2}
\end{equation}
In the present form, \eqref{eq:localSecondLaw1} holds for scalar fields $\Phi_a$. However, similar relations can be obtained for  fields in the vector, spinor or tensor representation.

A remark of caution is in order at this point. While the entropy current can be defined as in \eqref{eq:defEntropyCurrent} for situations that are very close to local equilibrium, it can be necessary to change this definition in situations where one deviates from equilibrium somewhat stronger. An example for this is relativistic fluid dynamics at second order in gradients, see for example ref.\ \cite{Israel:1979wp}. 

Note that the action $\Gamma_D[\Phi]$ that enters \eqref{eq:localSecondLaw1} and \eqref{eq:localSecondLaw2} yields precisely those terms in $\Gamma[\Phi]$ that contain the symbol $s_\text{R}({\cal L}_u)$ and account for dissipation.

\section{Example I: Scalar field with $O(N)$ symmetry}
\label{eq:ExampleScalarField}

\subsection{Dissipative equations of motion}

We discuss here the formalism developed in the previous sections for the example of a relativistic scalar field theory with global $O(N)$ symmetry. We take the analytic effective action to be of the form
\begin{equation}
\begin{split}
\Gamma[\varphi, g_{\mu\nu}, \beta^\mu]= \int d^d x \sqrt{g} {\bigg \{} & \frac{1}{2} Z(\rho,T) g^{\mu\nu} \partial_\mu \varphi_j  \partial_\nu \varphi_j + U(\rho,T) \\
& + \frac{1}{2} C(\rho,T) \, \left[ \varphi_j , s_\text{R}(u^\mu\partial_\mu) \right] \, \beta^\nu \partial_\nu \varphi_j
{\bigg \}},
\end{split}
\label{eq:AnalyticActionScalarField}
\end{equation}
where $\rho=\frac{1}{2}\varphi_j\varphi_j$ is the $O(N)$ invariant combination of fields, $T=1/\sqrt{-g_{\mu\nu}\beta^\mu \beta^\nu}$ is the temperature and $\beta^\mu=u^\mu/T$ is the ``inverse temperature vector''.
We first derive the equation of motion for $\varphi_j$ by variation according to the prescription in section \ref{sec:GeneralizedVariationalPrinciple}. The variation of \eqref{eq:AnalyticActionScalarField} at fixed metric $g_{\mu\nu}$ and inverse temperature field $\beta^\mu$ gives
\begin{equation}
\begin{split}
\delta\Gamma= \int d^d x \sqrt{g} {\bigg \{} & Z(\rho,T) g^{\mu\nu} \partial_\mu \delta \varphi_j  \partial_\nu \varphi_j + \frac{1}{2} Z^\prime(\rho,T) \varphi_m \delta\varphi_m \; g^{\mu\nu} \partial_\mu \varphi_j  \partial_\nu \varphi_j + U^\prime(\rho,T) \varphi_m \delta \varphi_m \\
& + \frac{1}{2} C(\rho,T) \, \left[\delta \varphi_j , s_\text{R}(u^\mu\partial_\mu) \right] \beta^\nu \partial_\nu \varphi_j
+ \frac{1}{2} C(\rho,T) \, \left[\varphi_j, s_\text{R}(u^\mu\partial_\mu)\right] \beta^\nu \partial_\nu \delta\varphi_j \\
& + \frac{1}{2} C^\prime(\rho,T) \varphi_m \delta\varphi_m \, \left[\varphi_j , s_\text{R}(u^\mu\partial_\mu) \right] \beta^\nu \partial_\nu \varphi_j
{\bigg \}}.
\end{split}
\label{eq:AnalyticActionScalarFieldVariation}
\end{equation}
The primes denote here derivatives with respect to $\rho$. The terms in the first line are standard variations of terms without discontinuities. 
For the first term in the second line of \eqref{eq:AnalyticActionScalarFieldVariation}, the commutator $[\delta \varphi_j, s_\text{R}(u^\mu\partial_\mu)] =\delta \varphi_j s_\text{R}(u^\mu\partial_\mu) -s_\text{R}(u^\mu\partial_\mu) \delta \varphi_j $ contains one term where the variation $\delta \varphi$ is to the left of the operator $s_\text{R}(u^\mu \partial_\mu)$ and for the retarded variation one has to set $s_\text{R}(u^\mu \partial_\mu)\to 1$ there. In contrast, for the second term, $\delta \varphi$ is to the right and one has to set $s_\text{R}(u^\mu \partial_\mu)\to -1$. In total, this results in $[\delta \varphi_j, s_\text{R}(u^\mu\partial_\mu)] \to 2 \delta \varphi_j$.

For the second term in the second line of \eqref{eq:AnalyticActionScalarFieldVariation} as well as for the term in the third line, the field variation $\delta \varphi_j$ is always to the right or to the left of $s_\text{R}(u^\mu \partial_\mu)$. Accordingly the commutator leads to a vanishing contribution by these terms. One can then transfer the remaining partial derivatives acting on field variations $\delta \varphi$ to other terms by partial integration and finds for the field equation \eqref{eq:FieldEqRet} at vanishing source $J=0$,
\begin{equation}
\begin{split}
& -\nabla_\mu \left[ Z(\rho,T) \partial^\mu \varphi_j \right] + \frac{1}{2} Z^\prime(\rho,T) \varphi_j \partial_\mu \varphi_m \partial^\mu \varphi_m +   U^\prime(\rho,T) \varphi_j  + C(\rho,T) \beta^\mu \partial_\mu \varphi_j  = 0.
\end{split}
\end{equation}
One can see this as a generalization of the Klein-Gordon equation with an additional damping term.

Note that for $N\geq 2$ the action \eqref{eq:AnalyticActionScalarField} has continuous global symmetries and associated conserved currents. Dissipation leads then to local equilibrium states with non-vanishing chemical potentials corresponding to the additional conserved quantum numbers. For a consistent description, these chemical potentials should actually be included in \eqref{eq:AnalyticActionScalarField} as additional parameter fields. Differential equations that determine their space-time evolution can be derived from the conservation laws. This will be discussed in more detail elsewhere.

\subsection{Energy-momentum tensor and entropy production}

The expectation value of the energy-momentum tensor follows from retarded variation of \eqref{eq:AnalyticActionScalarField} with respect to $g_{\mu\nu}$ at fixed $\varphi$ and $\beta^\mu$
\begin{equation}
\begin{split}
\langle T^{\mu\nu}(x) \rangle = & - \frac{2}{\sqrt{g}(x)} \frac{\delta \Gamma[\varphi,g_{\mu\nu},\beta^\mu]}{\delta g_{\mu\nu}}{\bigg |}_\text{ret} \\
= & Z(\rho,T) \partial^\mu \varphi_j \partial^\nu \varphi_j - \left(g^{\mu\nu}+u^\mu u^\nu T \frac{\partial}{\partial T}\right) \left\{ \frac{1}{2} Z(\rho,T) g^{\mu\nu} \partial_\mu \varphi_j  \partial_\nu \varphi_j + U(\rho,T) \right\}
\end{split}
\label{eq:enmomtensorScalarField}
\end{equation}
For the action \eqref{eq:AnalyticActionScalarField}, there is actually no contribution from the dissipative terms to the energy momentum tensor so that $\langle T^{\mu\nu}(x) \rangle = (\bar T_R)^{\mu\nu}(x)$. 
Note that the expression in \eqref{eq:enmomtensorScalarField} reduces to the standard form of the energy-momentum tensor for a scalar field when the temperature dependence drops out, as well as to the energy-momentum tensor for an ideal fluid with pressure $p=-U$ and enthalpy density $\epsilon + p= s T = -T\frac{\partial}{\partial T}U$ for $Z(\rho,T)=0$. 

The reduced action corresponding to \eqref{eq:AnalyticActionScalarField} is given by 
\begin{equation}
\Gamma_R[\varphi, g_{\mu\nu}, \beta^\mu]= \int d^d x \sqrt{g} {\bigg \{} \frac{1}{2} Z(\rho,T) g^{\mu\nu} \partial_\mu \varphi_j  \partial_\nu \varphi_j + U(\rho,T) {\bigg \}},
\label{eq:AnalyticActionScalarFieldReduced}
\end{equation}
and the dissipative part by
\begin{equation}
\Gamma_D[\varphi, g_{\mu\nu}, \beta^\mu]=  \int d^d x \sqrt{g} {\bigg \{}\frac{1}{2} C(\rho,T) \, \left[ \varphi_j , s_\text{R}(u^\mu\partial_\mu) \right] \, \beta^\nu \partial_\nu \varphi_j {\bigg \}}.
\label{eq:AnalyticActionScalarFieldDissipative}
\end{equation}
Accordingly, the vector field $K_\mu$ as defined in \eqref{eq:defKmu} is given by 
\begin{equation}
K_\mu = T^2 \beta_\mu T \frac{\partial}{\partial T} \left\{ \frac{1}{2} Z(\rho,T) g^{\alpha\beta}\partial_\alpha \varphi_j  \partial_\beta \varphi_j + U(\rho,T) \right\},
\end{equation}
and the analog of \eqref{eq:defMa} is in the present situation
\begin{equation}
\begin{split}
M_j = \frac{1}{\sqrt{g}} \frac{\delta \Gamma_D}{\delta \varphi_j} {\bigg |}_\text{ret} = & C(\rho, T) \beta^\mu \partial_\mu \varphi_j .
\end{split}
\label{eq:MjScalar}
\end{equation}
These quantities can be used in the analog of eq.\ \eqref{eq:520} in order to obtain an equation that fixes the space-time evolution of $\beta^\mu$. Particularly interesting is the entropy production. In the present context, one has
\begin{equation}
s^\mu = \beta^\mu \beta^\lambda K_\lambda =  - \beta^\mu \, T \frac{\partial}{\partial T} \left\{ \frac{1}{2} Z(\rho,T) g^{\alpha\beta}\partial_\alpha \varphi_j  \partial_\beta \varphi_j + U(\rho,T) \right\} ,
\end{equation}
and accordingly
\begin{equation}
s_G = -\frac{\partial}{\partial T} \left\{ \frac{1}{2}Z(\rho,T) g^{\alpha\beta}\partial_\alpha \varphi_j  \partial_\beta \varphi_j + U(\rho,T) \right\}
\end{equation}
can be seen as a generalized entropy density. Equation \eqref{eq:ent1} implies
\begin{equation}
\begin{split}
\nabla_\mu s^\mu = & C(\rho,T)  \left(\beta^\mu \partial_\mu \varphi_j \right) \left(\beta^\nu \partial_\nu \varphi_j \right) .\end{split}
\end{equation}
This is positive semi-definite indeed for 
\begin{equation}
C(\rho,T) \geq 0.
\end{equation}
This relation should therefore be seen as a requirement of the second law of thermodynamics in the present context. For the special case where the fluid is at rest $u^\mu=(1,0,0,0)$, entropy production becomes
\begin{equation}
\nabla_\mu s^\mu = \dot s_G = \frac{C(\rho,T)}{T^2} \dot \varphi_j \dot \varphi_j.
\end{equation}
The entropy increases when the field $\varphi_j$ oscillates. An example for such a situation is the period of reheating after inflation.

\section{Example II: Relativistic fluid dynamics}
\label{sec:FluidDynamics}

We discuss now how the equations of relativistic fluid dynamics can emerge from the formalism developed is sections \ref{sec:AnalyticQuantumEffectiveAction} - \ref{sec:EnergyMomentumConservationEntropyProduction}. We consider here a fluid without any conserved charges apart from energy and momentum. 

First, consider an effective action of the form
\begin{equation}
\Gamma[g_{\mu\nu}, \beta^\mu] = \Gamma_R[g_{\mu\nu}, \beta^\mu]= \int d^d x \sqrt{g} \;  U(T) ,
\end{equation}
where the effective potential $U(T) = - p(T)$ equals the negative of pressure, $T=1/\sqrt{-g_{\mu\nu}\beta^\mu \beta^\nu}$ is the temperature and $\beta^\mu=u^\mu/T$ is a combination of fluid velocity and temperature. The energy-momentum tensor obtained by variation with respect to $g_{\mu\nu}$ at fixed $\beta^\mu$ according to \eqref{eq:ExpValTmunu0} gives
\begin{equation}
T^{\mu\nu} = (\epsilon + p) u^\mu u^\nu + p g^{\mu\nu} ,
\end{equation}
where $\epsilon+p=T s = T \frac{\partial}{\partial T} p$ is the enthalpy density. This is the energy-momentum tensor of an ideal fluid. The space-time evolution of $\beta^\mu$ follows as usual from the covariant conservation law $\nabla_\mu T^{\mu\nu}=0$ or, equivalently, from general covariance as discussed in sect.\ \ref{sec:EnergyMomentumConservation}. That leads to the ideal relativistic fluid equations
\begin{equation}
u^\mu\partial_\mu \epsilon + (\epsilon+p) \nabla_\mu u^\mu = 0, \quad\quad\quad
(\epsilon + p) u^\mu \nabla_\mu u^\nu + \Delta^{\nu\mu}\partial_\mu p = 0,
\end{equation}
as usual. No dissipative effects are present for the ideal fluid and the entropy current $s^\mu = s u^\mu$ is conserved.

Let us now attempt to introduce shear and bulk viscous dissipation. The energy-momentum tensor must contain additional terms, the shear stress and bulk viscous pressure which should follow from a retarded variational derivative with respect to the metric. Consider the analytically continued effective action
\begin{equation}
\Gamma[g_{\mu\nu}, \beta^\mu] = \int d^d x \sqrt{g} \; \left\{ U(T) + \frac{1}{4}\left[ g_{\mu\nu}, s_\text{R}({\cal L}_u) \right] (2 \eta(T) \sigma^{\mu\nu} + \zeta(T) \Delta^{\mu\nu} \nabla_\rho u^\rho) \right\},
\label{eq:actionHydro}
\end{equation}
where ${\cal L}_u$ is the Lie derivative in the direction $u^\mu$, $\Delta^{\mu\nu} = u^\mu u^\nu + g^{\mu\nu}$ is a projector orthogonal to the fluid velocity $u^\mu=T \beta^\mu$, and
\begin{equation}
\sigma^{\mu\nu} = \left(\frac{1}{2} \Delta^{\mu\alpha} \Delta^{\mu\beta} +\frac{1}{2} \Delta^{\mu\beta} \Delta^{\mu\alpha} - \frac{1}{d-1} \Delta^{\mu\nu} \Delta^{\alpha\beta}\right) \nabla_\alpha u_\beta
\end{equation}
is a combination of covariant derivatives of the fluid velocity that is symmetric, transverse to the fluid velocity $u_\mu \sigma^{\mu\nu} = 0$, and trace-less $g_{\mu\nu} \sigma^{\mu\nu}=0$. The function $\eta(T)$ is actually the shear viscosity and $\zeta(T)$ is the bulk viscosity. 

The retarded functional derivative with respect to $g_{\mu\nu}$ of the action \eqref{eq:actionHydro} according to \eqref{eq:ExpValTmunu0} gives
\begin{equation}
\langle T^{\mu\nu} \rangle = (\epsilon + p) u^\mu u^\nu + p g^{\mu\nu} - 2 \eta \sigma^{\mu\nu} -\zeta \Delta^{\mu\nu} \nabla_\rho u^\rho,
\end{equation}
which is the energy-momentum tensor of a fluid with shear and bulk viscosity in the so-called first order formalism. It can be decomposed into the reduced part $(\bar T_R)^{\mu\nu} = (\epsilon + p) u^\mu u^\nu + p g^{\mu\nu}$ and the dissipative part $(\bar T_D)^{\mu\nu}= - 2 \eta \sigma^{\mu\nu} -\zeta \Delta^{\mu\nu} \nabla_\rho u^\rho$. The entropy production follows from \eqref{eq:ent1} as
\begin{equation}
\nabla_\mu s^\mu = \beta_\nu \nabla_\mu (\bar T_D)^{\mu\nu} = - (\nabla_\mu \beta_\nu)  (\bar T_D)^{\mu\nu} = \frac{1}{T}\left[2 \eta \sigma_{\mu\nu} \sigma^{\mu\nu} + \zeta (\nabla_\rho u^\rho)^2 \right],
\label{eq:entropyProductionNavierStokes}
\end{equation}
where we have used $\beta_\nu(\bar T_D)^{\mu\nu}=0$ and other symmetry properties. Eq.\ \eqref{eq:entropyProductionNavierStokes} is indeed the correct expression for entropy production of a relativistic fluid without charges in the first order formalism. The right hand side is positive semi-definite for $\eta\geq0$ and $\zeta\geq 0$.

One should keep in mind, that the first order approximation to viscous relativistic fluid dynamics has problems with causality and linear stability \cite{Israel:1979wp, Hiscock:1985zz}. In any case, the first order viscous fluid equations have physical significance only for very large length and time scales. For shorter lengths and times, higher order gradients become relevant. We postpone a discussion of second order viscous fluid dynamics, as well as a more detailed discussion of the first order approximation, to the future.

\section{Conclusions}
\label{sec:Conclusions}

We have discussed here how one can obtain causal, dissipative equations of motion from the analytically continued, one-particle irreducible quantum effective action (or analytic effective action for short). The latter has first been defined in the Euclidean domain for situations that can be described by a generalized Gibbs ensemble in local thermal equilibrium with a temperature and fluid velocity that depend on space and time. By studying the analytic properties of correlation functions, first in the linear response regime but subsequently also in more general, non-linear situations, it was possible to gain some insights into the general analytic structure of correlation functions and thereby the analytic structure of the quantum effective action itself.

The most important element is the branch cut behavior that can arise for example in the two-point correlation functions along the real frequency axis. In a local equilibrium situation and in position space, this branch cut can be parametrized in terms of the symbol
\begin{equation}
s_\text{R}({\cal L}_u) = s_\text{R}(u^\mu \partial_\mu) = \text{sign}\left( \text{Re} \left( u^\mu(x) \tfrac{\partial}{\partial x^\mu} \right) \right) .
\label{eq:sRConclusions}
\end{equation}
This is the formal analog of $s_\text{I}(\omega) = \text{sign}(\text{Im} (\omega))$, the sign function that allows to parametrize the branch cut in the frequency domain.
The vector field $u^\mu(x)$, that enters \eqref{eq:sRConclusions} denotes the fluid velocity corresponding to the generalized local Gibbs ensemble.

Based on the symbol \eqref{eq:sRConclusions} one can formulate possible structures that can arise in the analytic effective action $\Gamma[\Phi]$. It is actually convenient to formulate the latter directly in general coordinates with a metric $g_{\mu\nu}(x)$ and for a generalized Gibbs ensemble described by the combination of fluid velocity and temperature $\beta^\mu(x)= u^\mu(x) / T(x)$. One can then also formulate a generalized variational principle that leads to causal and dissipative equations of motion by formulating rules how the sign \eqref{eq:sRConclusions} has to be chosen, depending on whether the field variation $\delta \Phi$ is to the left or to the right of this formal operator. That the resulting equations of motion are causal can be seen on general grounds in flat space, see eqs.\ \eqref{eq:42} and \eqref{eq:Causality}. (The issue may be more involved in the context of general relativity with curved space.)

One can also discuss the important issue of energy-momentum conservation in the present formalism. To that end it is useful to work with a reduced action $\Gamma_R[\Phi]$ that is obtained from the full analytic effective action $\Gamma[\Phi]$ by dropping the discontinuous terms, i.\ e.\ by formally setting $s_\text{R}(u^\mu \partial_\mu) \to 0$. A very interesting relation follows from studying the implications of general coordinate invariance of the reduced action $\Gamma_R[\Phi,g_{\mu\nu},\beta^\mu]$. Besides the field expectation value $\Phi$, the reduced action depends also on the metric $g_{\mu\nu}$ and the temperature and fluid velocity field $\beta^\mu=u^\mu/T$. The condition that $\Gamma_R[\Phi,g_{\mu\nu},\beta	^\mu]$ be general covariant leads to four additional, non-trivial differential equations which can be used to fix the space-time evolution of temperature $T(x)$ and fluid velocity $u^\mu(x)$. (Alternatively, such relations can be derived from the covariant conservation law $\nabla_\mu T^{\mu\nu}$ and the equations of motion, similar to how it is usually done in the formalism of relativistic fluid dynamics.)

From the consequences of general covariance, one can in particular derive an equation that describes local entropy production. An inequality of the form $\nabla_\mu s^\mu(x)\geq 0$, where $s^\mu(x)$ is the entropy current, leads to a local form of the second law of thermodynamics and puts some constraints on the dissipative dynamics. 

All these relations can be studied in a more concrete form for the example of an effective action that describes a scalar field with $O(N)$ symmetry as is done in section \ref{eq:ExampleScalarField}. In particular, the dissipative equation of motion corresponds then to a generalized Klein-Gordon equation with a dissipative damping term. The entropy production inequality leads to a constraint for the form of the analytic action $\Gamma[\Phi]$. With this constraint taken into account, the entropy is non-decreasing, indeed. Entropy can be produced for example by an oscillating scalar field, a mechanism that plays a role for reheating after inflation in the early universe. 

The present paper had to leave open a couple of points that would deserve further attention. In particular, we have concentrated here of the dissipative equations of motion, but of course, the analytic effective action contains also very valuable information about various correlation functions. These correlation functions characterize fluctuations for a generalized local Gibbs ensemble and generalize directly both the thermal correlation functions as one can study them in a complete thermal equilibrium, for example in the Matsubara formalism, as well as correlation functions of quantum fields in vacuum. For a recent discussion of correlation functions of fluid dynamic variables see ref.\ \cite{Kovtun:2012rj}.

In the present paper we also restricted the discussion of relativistic fluid dynamics in section \ref{sec:FluidDynamics} to the first order approximation. This can be extended to higher orders in gradients as well as fluids with additional conserved charges and order parameter fields $\Phi_a$. This will be discussed in more detail in subsequent publications.

The reader may also wonder how the formalism discussed here relates to more general out-of-equilibrium formalisms and in particular the Schwinger-Keldysh closed time path formalism. Some discussion of this issue can be found in the introduction but a few points may actually demand further clarifications to be given elsewhere. The relation may also become more clear once more detailed applications of the present formalism have been worked out. 

Finally, an interesting question is also how one can determine the analytic effective action from microscopic calculations. In principle, many different possibilities exist, ranging from perturbation theory to non-perturbative numerical methods, AdS/CFT based approaches or the functional renormalization group\footnote{See ref.\ \cite{Floerchinger:2011sc} for an approach using analytically continued renormalization group flow equations.}. We believe that the insights gained here into the generic structure of $\Gamma[\Phi]$ are helpful for all these approaches.

We are optimistic that the formalism based on the analytic effective action, as it has been developed here, has many interesting applications in various fields of physics (and perhaps even chemistry). This could range from the smallest scales as they are probed by relativistic heavy ion collisions up to the largest scales of cosmology. 

\begin{appendix}

\section{Two-point functions in linear response theory}
\label{sec:appA}

In this appendix we recall some standard definitions and results about different two-point correlation functions for a quantum field theory in thermal equilibrium in the linear response regime. In contrast to the other sections of this paper it will be useful here to work in the operator representation of quantum field theory but the results can be generalized easily to the functional integral representation. In a slight abuse of notation we will denote operators and fields (in the functional integral formalism) by the same symbol but it should always be clear from the context what is meant. 

Let us start from the correlation function of two operators $\phi_a$ and $\phi_b$ which might be elementary field operators but could as well be composite operators such as pairing fields or for example energy density, 
\begin{equation}
\Delta_{\phi_a\phi_b}^+(x-y) = \langle \phi_a(x) \phi_b(y) \rangle = \text{tr} \left\{ \rho \, \phi_a(x) \phi_b(y) \right\} .
\end{equation}
For some purposes, the above notation where the fields $\phi_a$ and $\phi_b$ appear in the index of the function $\Delta^+_{\phi_a\phi_b}$ is useful, while for others the simplified notation
\begin{equation}
\Delta_{ab}^+(x-y) = \Delta_{\phi_a\phi_b}^+(x-y)
\end{equation}
has advantages. In this appendix, we will take the liberty to use both notations in parallel while we stick to the simplified notation in the main text.

We take the density matrix to correspond to a thermal equilibrium with temperature $T$ and fluid velocity $u^\mu$ (a possible chemical potential can be included as an external gauge field),
\begin{equation}
\rho = \frac{1}{Z} e^{\frac{u_\nu \mathscr{P}^\nu}{T}} .
\end{equation}
The operators $\phi_a$ and $\phi_b$ have the Heisenberg representation
\begin{equation}
\phi(x) = e^{-i x_\nu \mathscr{P}^\nu} \phi(0) e^{i x_\nu \mathscr{P}^\nu} .
\end{equation}
Introduce now a complete set of states which are eigenstates of the four-momentum operator $\mathscr{P}^\nu = (H, \vec P)$ ,
\begin{equation}
\mathscr{P}^\nu | m \rangle = p^\nu_m |m \rangle.
\end{equation}
The normalization is taken to be such that
\begin{equation}
\mathbbm{1} = | m \rangle \langle m |, \quad\quad\quad \langle m | n \rangle = \delta_{mn} .
\end{equation}
One obtains
\begin{equation}
\Delta_{\phi_a\phi_b}^+(x-y) =\sum_{m,l} \frac{1}{Z} e^{\frac{u_\nu p^\nu_m }{T}} e^{i (p_l - p_m)(x-y)} \langle m | \phi_a(0) | l \rangle \langle l | \phi_b(0) | m \rangle .
\end{equation}
Introduce also the momentum space representation
\begin{equation}
\Delta^+_{\phi_a\phi_b}(x-y) = \int \frac{d^4 p}{(2\pi)^4} e^{ip(x-y)} \; \Delta^+_{\phi_a\phi_b}(p) ,
\end{equation}
with 
\begin{equation}
\Delta_{\phi_a\phi_b}^+(p) =\sum_{m,l} \delta^{(4)}(p-p_l+p_m)  \frac{1}{Z} e^{\frac{u_\nu p^\nu_m}{T}} \langle m | \phi_a(0) | l \rangle \langle l | \phi_b(0) | m \rangle .
\end{equation}
Note that for $T \to 0$ only the ground state survives in the sum over $m$ and one has
\begin{equation}
\Delta_{\phi_a\phi_b}^+(p) =\sum_l \delta^{(4)}(p-p_l) \langle 0 | \phi_a(0) | l \rangle \langle l | \phi_b(0) | 0 \rangle  \quad\quad\quad\quad\quad (T \to 0).
\label{eq:TzeroLimit}
\end{equation}
In the case where $\phi_a$ and $\phi_b$ are elementary particle operators, the sum in \eqref{eq:TzeroLimit} contains positive frequencies, $p^0_l \geq 0$, only. This is actually the reason for the name $\Delta^+$. For non-vanishing temperature this is not the case, however. 
One has also
\begin{equation}
\begin{split}
\langle \phi_b(y) \phi_a(x) \rangle = & \Delta_{\phi_b\phi_a}^+(y-x) 
= \int \frac{d^4 p}{(2\pi)^4} e^{ip(x-y)} \Delta_{\phi_b\phi_a}^+(-p)
\end{split}
\end{equation}
with
\begin{equation}
\Delta_{\phi_b\phi_a}^+(-p) = e^{\frac{u_\nu p^\nu}{T}}\Delta_{\phi_a\phi_b}^+(p) .
\end{equation}
Define now also the spectral and statistical correlation functions by
\begin{equation}
\begin{split}
\Delta^\rho_{\phi_a\phi_b}(x-y) = & \langle \left[\phi_a(x), \phi_b(y) \right]_\mp \rangle = \text{tr} \left\{ \rho \left[\phi_a(x), \phi_b(y)\right]_\mp \right\} = \int_p e^{ip(x-y)} \Delta^\rho_{\phi_a\phi_b}(p), \\
\Delta^S_{\phi_a\phi_b}(x-y) = &\frac{1}{2} \langle \left[\phi_a(x), \phi_b(y) \right]_\pm \rangle = \frac{1}{2} \text{tr} \left\{ \rho \left[\phi_a(x), \phi_b(y)\right]_\pm \right\} = \int_p e^{ip(x-y)} \Delta^S_{\phi_a\phi_b}(p).
\end{split}
\label{eq:defDeltaRhoDeltaS}
\end{equation}
If at least one of the operators $\phi_a$ or $\phi_b$ is bosonic, the spectral function $\Delta^\rho$ involves the commutator (upper sign), if both $\phi_a$ and $\phi_b$ are fermionic the anti-commutator (lower sign). For the statistical propagator the situation is reversed. One has 
\begin{equation}
\begin{split}
\Delta^\rho_{\phi_a\phi_b}(p) = & \Delta_{\phi_a\phi_b}^+(p) \mp \Delta_{\phi_b\phi_a}^+(-p) = \left( 1 \mp e^{\frac{u_\nu p^\nu}{T}} \right) \Delta_{\phi_a\phi_b}^+(p) ,\\
\Delta^S_{\phi_a\phi_b}(p) = & \frac{1}{2} \left[\Delta_{\phi_a\phi_b}^+(p) \pm \Delta_{\phi_b\phi_a}^+(-p) \right] = \frac{1}{2} \left( 1 \pm e^{\frac{u_\nu p^\nu}{T}} \right) \Delta_{\phi_a\phi_b}^+(p) .
\end{split}
\end{equation}
This yields the so-called fluctuation-dissipation relation
\begin{equation}
\Delta^S_{\phi_a\phi_b}(p) = \left[ \frac{1}{2} \pm n_{B/F}(-u_\nu p^\nu) \right] \Delta^\rho_{\phi_a\phi_b}(p) ,
\label{eq:FluctDissRelation}
\end{equation}
where the Bose and Fermi occupation number functions are
\begin{equation}
n_{B/F}(\omega) = \frac{1}{e^{\frac{\omega}{T}}\mp 1}.
\end{equation}
Note that the square bracket in \eqref{eq:FluctDissRelation} is anti-symmetric under $p^\nu \to - p^\nu$.

One defines also the Feynman, retarded and advanced propagators by
\begin{equation}
\begin{split}
-i \Delta^F_{\phi_a\phi_b}(x-y) = & \langle T \, \phi_a(x) \phi_b(y) \rangle = \theta(x^0-y^0) \langle \phi_a(x) \phi_b(y) \rangle \pm \theta(y^0-x^0) \langle \phi_b(y) \phi_a(x) \rangle , \\
-i \Delta^R_{\phi_a\phi_b}(x-y) = & \theta(x^0 - y^0) \langle \left[ \phi_a(x) , \phi_b(y)\right]_\mp \rangle , 
\\
-i \Delta^A_{\phi_a\phi_b}(x-y) = & - \theta(y^0 - x^0) \langle \left[ \phi_a(x) , \phi_b(y)\right]_\mp \rangle , 
\end{split}
\label{eq:DefFeynmanRetardedAdvanced}
\end{equation}
with corresponding momentum space representations $\Delta^F_{\phi_a\phi_b}(p)$, $\Delta^R_{\phi_a\phi_b}(p)$ and $\Delta^A_{\phi_a\phi_b}(p)$. From the Feynman propagator one obtains via analytic continuation the Matsubara propagator. In momentum space ($\Delta^\mu_{\;\;\nu} = u^\mu u_\nu + \delta^\mu_{\;\;\nu}$ is the projector orthogonal to the fluid velocity),
\begin{equation}
\begin{split}
& \Delta^M(i \omega_n, \Delta^\mu_{\;\;\nu} p^\nu) = \Delta^F(-u_\nu p^\nu = i \omega_n , \Delta^\mu_{\;\;\nu} p^\nu) .
\end{split}
\end{equation}
The following relations follow directly from the definitions
\begin{equation}
\Delta^R_{\phi_a \phi_b}(x-y) = \pm\Delta^A_{\phi_b \phi_a}(y-x),  \quad\quad\quad \Delta^A_{\phi_a \phi_b}(x-y) = \pm \Delta^R_{\phi_b \phi_a}(y-x) ,
\end{equation}
or, in momentum space,
\begin{equation}
\Delta^R_{\phi_a \phi_b}(p) = \pm \Delta^A_{\phi_b \phi_a}(-p),  \quad\quad\quad \Delta^A_{\phi_a \phi_b}(p) = \pm \Delta^R_{\phi_b \phi_a}(-p) .
\end{equation}
One has also
\begin{equation}
    \Delta^R_{\phi_a \phi_b}(x-y) = \pm \Delta^{R*}_{\phi_a^\dagger \phi_b^\dagger}(x-y) , \quad\quad\quad \Delta^A_{\phi_a \phi_b}(x-y) = \pm \Delta^{A*}_{\phi_a^\dagger \phi_b^\dagger}(x-y) ,
\end{equation}
or, in momentum space,
\begin{equation}
\Delta^R_{\phi_a \phi_b}(p) = \pm \Delta^{R*}_{\phi_a^\dagger \phi_b^\dagger}(-p) , \quad\quad\quad \Delta^A_{\phi_a \phi_b}(p) = \pm \Delta^{A*}_{\phi_a^\dagger \phi_b^\dagger}(-p) .
\end{equation}

One can also discuss different discrete symmetries. Particularly useful is time reversal. Assume that the field operators transform as
\begin{equation}
\textsf{T} \phi_a(x^0,\vec x) \textsf{T}^{-1} = \phi_{a\textsf{T}}(-x^0,\vec x)
\end{equation}
and that the density matrix is invariant up to a time-reversal breaking external parameter $B$ (such as a magnetic field)
\begin{equation}
\textsf{T} \rho(B) \textsf{T}^{-1} = \rho(-B) .
\end{equation}
One can show that the following relations are fulfilled (Onsager 1931)
\begin{equation}
\begin{split}
\Delta^R_{\phi_a\phi_b}(x^0-y^0,\vec x-\vec y; B) =  \Delta^R_{\phi_{b \textsf{T}}^\dagger \phi_{a \textsf{T}}^\dagger}(x^0 - y^0, -\vec x + \vec y; - B) , \\
\Delta^A_{\phi_a\phi_b}(x^0-y^0, \vec x - \vec y; B) =  \Delta^A_{\phi_{b \textsf{T}}^\dagger \phi_{a \textsf{T}}^\dagger}(x^0 - y^0, -\vec x + \vec y; -B) ,
\end{split}
\end{equation}
or, in momentum space,
\begin{equation}
\begin{split}
\Delta^R_{\phi_a\phi_b}(p^0,\vec p; B) =  \Delta^R_{\phi_{b \textsf{T}}^\dagger \phi_{a \textsf{T}}^\dagger}(p^0, -\vec p ; - B) , \\
\Delta^A_{\phi_a\phi_b}(p^0, \vec p; B) =  \Delta^A_{\phi_{b \textsf{T}}^\dagger \phi_{a \textsf{T}}^\dagger}(p^0, -\vec p; -B) .
\end{split}
\end{equation}
In particular, for hermitean fields with definite time reversal parity such that $\phi_{a\textsf{T}}^\dagger(x) = \eta_a \phi_a(x)$, one has
\begin{equation}
\Delta^R_{\phi_a\phi_b}(p^0,\vec p; B) =  \eta_a \eta_b \;  \Delta^R_{\phi_b\phi_a}(p^0, -\vec p ; - B) ,
\end{equation}
and similar for the advanced correlation function.

\begin{figure}
\centering
\includegraphics[width=0.5\textwidth]{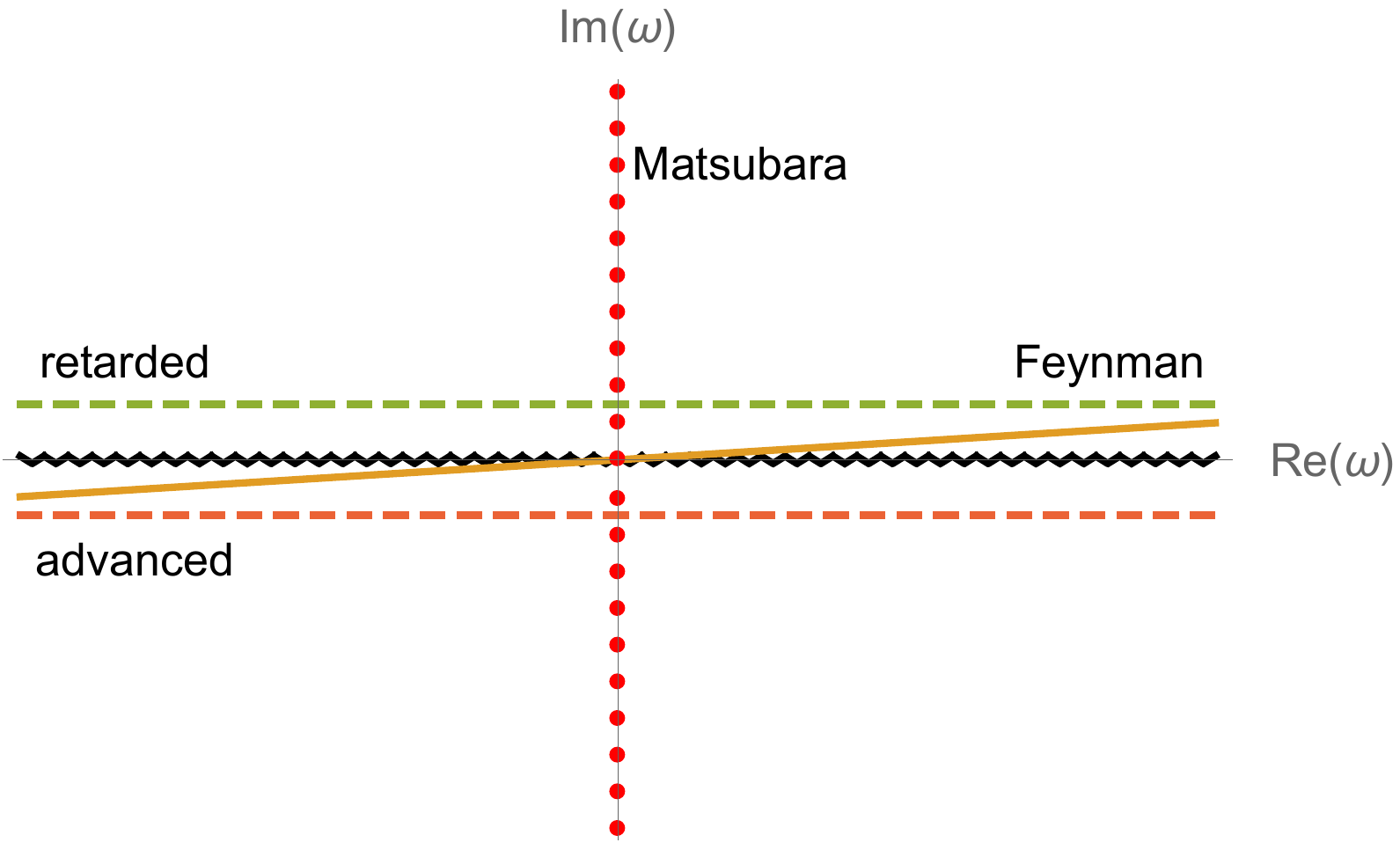}
\caption{Contours for obtaining different propagators from the complex argument function $G(p)$.}
\label{fig1}
\end{figure}

The spectral correlation function is directly related to the spectral density
\begin{equation}
\Delta^\rho_{\phi_a\phi_b}(p)=  2\pi \; \rho_{\phi_a\phi_b}\left(-p^2, -u_\nu p^\nu \right) .
\label{eq:DefSpectralDensity}
\end{equation}
At non-zero temperature it depends both on $p^2=-(p^0)^2+\vec p^2$ and the frequency in the fluid rest frame $-u_\nu p^\nu$. It depends also on the temperature $T$ as well as on external fields such as a chemical potential. In the zero-temperature limit one recovers the conventional vacuum spectral density as
\begin{equation}
\lim_{T \to 0} \rho_{\phi_a\phi_b} \left(-p^2, - u_\nu p^\nu\right) = \text{sign}(-u_\nu p^\nu) \, \rho_{\phi_a\phi_b}(-p^2).
\label{eq:vacuumLimitRho}
\end{equation}
Define now the complex argument Greens function by
\begin{equation}
G_{\phi_a\phi_b}\left(-u_\nu p^\nu, \Delta^\mu_{\;\;\nu} p^\nu \right) = \int_{- \infty}^\infty dw \; \rho_{\phi_a\phi_b}\left(w^2- \Delta_{\mu\nu} p^\mu p^\nu, w \right) \frac{1}{w+u_\nu p^\nu}.
\label{eq:GSpectralRep}
\end{equation}
The integral over $w$ is along the real axis. The function $G$ can be evaluated for $\omega = - u_\nu p^\nu \in \mathbbm{C}$. It has a brach cut or poles along the real $\omega$ axis. One can show that one obtains the Feynman, retarded, advanced and Matsubara Greens functions by evaluating $G$ on the contours shown in Fig.\ \ref{fig1}, shifted slightly away from the real axis. 
More explicit,
\begin{equation}
\begin{split}
\Delta^R_{\phi_a\phi_b}(p) = & G_{\phi_a\phi_b}\left(-u_\nu p^\nu + i \epsilon, \Delta^\mu_{\;\;\nu} p^\nu \right) , \\
\Delta^A_{\phi_a\phi_b}(p) = & G_{\phi_a\phi_b}\left(-u_\nu p^\nu - i \epsilon, \Delta^\mu_{\;\;\nu} p^\nu \right) , \\
\Delta^F_{\phi_a\phi_b}(p) = & G_{\phi_a\phi_b}\left(-u_\nu p^\nu + i \epsilon\;  \text{sign}(-u_\nu p^\nu), \Delta^\mu_{\;\;\nu} p^\nu \right) , \\
\Delta^M_{\phi_a\phi_b}(p) = & G_{\phi_a\phi_b}\left(i \omega_n , \Delta^\mu_{\;\;\nu} p^\nu \right) .
\end{split}
\label{eq:RetAdvFeynMatFromCompArgGreensFunc}
\end{equation}
The spectral correlation function can be obtained from
\begin{equation}
\Delta^\rho_{\phi_a\phi_b}(p) = -i \Delta^R_{\phi_a\phi_b}(p) + i \Delta^A_{\phi_a\phi_b}(p) ,
\label{eq:rhoRA}
\end{equation}
an identity that follows directly from the definitions \eqref{eq:defDeltaRhoDeltaS} and \eqref{eq:DefFeynmanRetardedAdvanced}. Finally, the statistical correlation function can be obtained from this via the fluctuation-dissipation relation \eqref{eq:FluctDissRelation}.
In the vacuum limit $T\to0$ one can use \eqref{eq:vacuumLimitRho} to rewrite \eqref{eq:GSpectralRep} in the standard form
\begin{equation}
G_{\phi_a\phi_b}(p) = \int_0^\infty d\mu^2 \;\rho_{\phi_a\phi_b}(\mu^2) \; \frac{1}{p^2 + \mu^2} .
\end{equation}

As is discussed in more detail in the main text, the function $G_{\phi_a\phi_b}(p)$ is obtained from the second functional derivative of the Schwinger functional $W[J]$. Similarly, its inverse
\begin{equation}
P_{\phi_a\phi_b}(p) = G_{\phi_a\phi_b}^{-1}(p) ,
\end{equation}
is obtained from the second functional derivative of the effective action $\Gamma[\phi]$. Similar to $G_{\phi_a\phi_b}(p)$, the function $P_{\phi_a\phi_b}(p)$ (or more specific its eigenvalues) might have brach cuts and zero-crossings along the axis of real $\omega=-u_\nu p^\nu$ but nowhere else.

One can decompose the inverse complex-argument two-point function
\begin{equation}
P_{\phi_a \phi_b}(p) = P_{1,\phi_a\phi_b}(p) - i s_\text{I}(-u_\nu p^\nu) \, P_{2,\phi_a\phi_b}(p) ,
\label{eq:DecomposeInverseComplexArgGF}
\end{equation}
where $s_\text{I}(\omega) = \text{sign}(\text{Im} \; \omega)$. Both functions $P_{1,\phi_a \phi_b}(p)$ and $P_{2,\phi_a \phi_b}(p)$ are regular when crossing the real frequency axis. However, the sign $s_\text{I}(-u_\nu p^\nu)$ changes, which leads to a brach cut behavior for the function $P_{\phi_a \phi_b}(p)$. The function $P_{2,\phi_a \phi_b}(p)$ parametrizes the strength of the branch cut.

For special cases of the operators $\phi_a$, $\phi_b$ one can make further going statements. In particular for $\phi_b(x) = \phi_a^*(x)$ one has $\rho(-p^2,-u_\nu p^\nu) \in \mathbbm{R}$ with corresponding relations for the different correlation functions. If the operator $\phi_b$ corresponds to the conjugate momentum of the operator $\phi_a$, i. e. $\phi_b(x) = - i\Pi_{\phi_a}(x)$ such that they fulfill a canonical commutation relation at equal time
\begin{equation}
[\phi_a(t,\vec x), \phi_b(t,\vec y)] = \delta^{(3)}(\vec x-\vec y) ,
\end{equation}
one has
\begin{equation}
\int_{-\infty}^\infty d p^0 \; \rho_{\phi_a \phi_b}((p^0)^2 -\vec p^2, p^0 ) = 1,
\end{equation}
for all values of $\vec p$.

\end{appendix}

\section*{Acknowledgements}
The author would like to thank J.~Berges, A.~Kurkela, M.~M.~Scherer, N.~Tetradis, R.~Venugopalan, C.~Wetterich and U. A.~Wiedemann for interesting discussions.

\end{document}